\documentclass[longauth]{aa}  

\usepackage{graphicx}
\usepackage{float}
\usepackage{hyperref}
\usepackage{txfonts}
\begin{document}

   \title{Precision measurement of a brown dwarf mass in a binary system in the microlensing event OGLE-2019-BLG-0033/MOA-2019-BLG-035}

   \author{A. Herald \inst{\ref{Padova}}
          \and
          A. Udalski\inst{\ref{O1}}
          V. Bozza\inst{\ref{Salerno},\ref{INFN}}
          \and
          P.Rota\inst{\ref{Salerno},\ref{INFN}}
         \and
           I.A. Bond\inst{\ref{K4}} \and
          J.C. Yee\inst{\ref{S6}}\and
           S. Sajadian\inst{\ref{M1}}
           \\ and \\ \vspace{0.2cm}
           P. Mr{\'o}z\inst{\ref{O1}} \and
           R. Poleski\inst{\ref{O1}} \and
           J. Skowron\inst{\ref{O1}} \and
           M.K. Szyma{\'n}ski\inst{\ref{O1}} \and
           I. Soszy{\'n}ski\inst{\ref{O1}} \and
           P. Pietrukowicz\inst{\ref{O1}} \and
           S. Koz{\l}owski\inst{\ref{O1}} \and
           K. Ulaczyk\inst{\ref{O2}} \and
           K.A. Rybicki\inst{\ref{O1}} \and
           P. Iwanek\inst{\ref{O1}} \and
           M. Wrona\inst{\ref{O1}} \and
           M. Gromadzki\inst{\ref{O1}}
            \\The OGLE collaboration\\ \vspace{0.2cm}
          F. Abe\inst{\ref{K1}} \and
          R. Barry\inst{\ref{K2}} \and
          D.P.~Bennett\inst{\ref{K2},\ref{K3}} \and
          A. Bhattacharya\inst{\ref{K2},\ref{K3}} \and
          A. Fukui\inst{\ref{K5},\ref{K6}} \and
          H. Fujii\inst{\ref{K1}} \and
          Y. Hirao\inst{\ref{K7}} \and
          Y. Itow\inst{\ref{K1}} \and
          R. Kirikawa\inst{\ref{K7}} \and
          I. Kondo\inst{\ref{K7}} \and
          N. Koshimoto\inst{\ref{K5}} \and
          Y. Matsubara \inst{\ref{K1}} \and
          S. Matsumoto \inst{\ref{K7}} \and
          S. Miyazaki \inst{\ref{K7}} \and
          Y. Muraki \inst{\ref{K1}} \and
          G. Olmschenk\inst{\ref{K2}} \and
          C. Ranc\inst{\ref{K8}} \and
          A. Okamura\inst{\ref{K7}} \and
          N.J. Rattenbury\inst{\ref{K9}} \and
          Y. Satoh\inst{\ref{K7}} \and
          T. Sumi\inst{\ref{K7}} \and
          D. Suzuki\inst{\ref{K7}} \and
          S. Ishitani Silva\inst{\ref{K10},\ref{K2}} \and
          T. Toda\inst{\ref{K7}} \and
          P.J. Tristram \inst{\ref{K11}} \and
          A. Vandorou\inst{\ref{K2},\ref{K3}}\and
          H. Yama\inst{\ref{K7}}
          \\ The MOA collaboration \\ \vspace{0.2cm}
          C.A. Beichman\inst{\ref{S1}}\and
          G. Bryden\inst{\ref{S2}}\and
          S. Calchi Novati\inst{\ref{S1}}\and
          S. Carey\inst{\ref{S1}}\and
          B.S. Gaudi\inst{\ref{S3}}\and
           A. Gould\inst{\ref{S4},\ref{S3}}\and
         C.B. Henderson\inst{\ref{S1}}\and
          S. Johnson\inst{\ref{S3}}\and
          Y. Shvartzvald\inst{\ref{S5}}\and
          W. Zhu\inst{\ref{L1}}
          \\ The {\it Spitzer} team \\ \vspace{0.2cm}
          M. Dominik\inst{\ref{M2}} \and
M. Hundertmark\inst{\ref{K8}}  \and
U. G. J{\o}rgensen\inst{\ref{M4}}  \and
P. Longa-Pe{\~n}a\inst{\ref{M5}}  \and
J. Skottfelt\inst{\ref{M6}}   \and 
J. Tregloan-Reed\inst{\ref{M7}}   \and
N. Bach-M{\o}ller\inst{\ref{M4}}   \and
M. Burgdorf\inst{\ref{M8}}   \and
G. D'Ago\inst{\ref{M9}}   \and
L. Haikala\inst{\ref{M7}}  \and
J. Hitchcock\inst{\ref{M2}}   \and
E. Khalouei\inst{\ref{M10}} \and
N. Peixinho\inst{\ref{M5}}   \and
S. Rahvar\inst{\ref{M10}}   \and
C. Snodgrass\inst{\ref{M11}}   \and
J. Southworth\inst{\ref{M12}}   \and
P. Spyratos\inst{\ref{M12}}   
          \\ The MiNDSTEp consortium \\  \vspace{0.2cm}
          W. Zang\inst{\ref{L1}} \and 
          H. Yang\inst{\ref{L1}} \and 
          S. Mao\inst{\ref{L1},\ref{L2}} \and
          E. Bachelet\inst{\ref{L3}} \and
          D. Maoz\inst{\ref{L5}} \and
          R.A. Street\inst{\ref{L3}} \and
          Y. Tsapras\inst{\ref{K8}} \and
          G.W. Christie\inst{\ref{U1}} \and
          T. Cooper\inst{\ref{U5}} \and
          L. de Almeida\inst{\ref{U2},\ref{U9}} \and
          J.-D. do Nascimento Jr\inst{\ref{U2},\ref{S6}} \and
          J. Green\inst{\ref{U5}} \and
          C. Han\inst{\ref{U10}} \and
          S. Hennerley\inst{\ref{U5}} \and
          A. Marmont\inst{\ref{U5}} \and
          J. McCormick\inst{\ref{U6}} \and
          L.A.G. Monard\inst{\ref{U7}} \and
          T. Natusch\inst{\ref{U1},\ref{U8}}\and
          R. Pogge\inst{\ref{S3}}
          \\ The LCO \& $\mu$FUN collaboration
          }

   \institute{Dipartimento di Fisica e Astronomia ``Galileo Galilei'', Universit\`a di Padova, Vicolo dell'Osservatorio 3, Padova I-35122, Italy \label{Padova}
   \and
   Astronomical Observatory, University of Warsaw, Al. Ujazdowskie 4, 00-478, Warszawa, Poland \label{O1}
   \and
   Dipartimento di Fisica ``E.R. Caianiello'', Universit\`a degli studi di Salerno, Via Giovanni Paolo II 132, I-84084 Fisciano (SA), Italy \label{Salerno}
   \and
   Istituto Nazionale di Fisica Nucleare, Sezione di Napoli, Via Cintia, 80126 Napoli (NA), Italy \label{INFN}
   \and
School of Mathematical and Computational Sciences, Massey University, Auckland 0745, New Zealand\label{K4}
\and
Center for Astrophysics $|$ Harvard \& Smithsonian, 60 Garden St.,Cambridge, MA 02138, USA\label{S6}
\and
    Department of Physics, Isfahan University of Technology, Isfahan, Iran \label{M1}
 \and
Department of Physics, University of Warwick, Gibbet Hill Road, Coventry, CV4~7AL,~UK \label{O2}
\and
Institute for Space-Earth Environmental Research, Nagoya University, Nagoya 464-8601, Japan\label{K1}
\and
Code 667, NASA Goddard Space Flight Center, Greenbelt, MD 20771, USA\label{K2}
\and
Department of Astronomy, University of Maryland, College Park, MD 20742, USA\label{K3}
\and
Department of Earth and Planetary Science, Graduate School of Science, The University of Tokyo, 7-3-1 Hongo, Bunkyo-ku, Tokyo 113-0033, Japan\label{K5}
\and
Instituto de Astrof\'isica de Canarias, V\'ia L\'actea s/n, E-38205 La Laguna, Tenerife, Spain\label{K6}
\and
Department of Earth and Space Science, Graduate School of Science, Osaka University, Toyonaka, Osaka 560-0043, Japan\label{K7}
\and
Zentrum f{\"u}r Astronomie der Universit{\"a}t Heidelberg, Astronomisches Rechen-Institut, M{\"o}nchhofstr.\ 12-14, 69120 Heidelberg, Germany\label{K8}
\and
Department of Physics, University of Auckland, Private Bag 92019, Auckland, New Zealand\label{K9}
\and
Department of Physics, The Catholic University of America, Washington, DC 20064, USA\label{K10}
\and
University of Canterbury Mt.\ John Observatory, P.O. Box 56, Lake Tekapo 8770, New Zealand\label{K11}
\and
IPAC, Mail Code 100-22, Caltech, 1200 E. California Blvd., Pasadena, CA 91125, USA\label{S1}
\and
Jet Propulsion Laboratory, California Institute of Technology, 4800 Oak Grove Drive, Pasadena, CA 91109, USA\label{S2}
\and
Department of Astronomy, Ohio State University, 140 W. 18th Ave., Columbus, OH  43210, USA\label{S3}
\and
Max-Planck-Institute for Astronomy, K\"onigstuhl 17, 69117 Heidelberg, Germany\label{S4}
\and
Department of Particle Physics and Astrophysics, Weizmann Institute of Science, Rehovot 76100, Israel\label{S5}
\and
Department of Astronomy, Tsinghua University, Beijing 100084, China \label{L1}
\and
 Centre for Exoplanet Science, SUPA, School of Physics \& Astronomy, University of St Andrews, North Haugh, St Andrews KY16 9SS, UK  \label{M2}
\and
Centre for ExoLife Sciences, Niels Bohr Institute, {\O}ster Voldgade 5, 1350 Copenhagen, Denmark  \label{M4}
\and
Unidad de Astronom{\'{\i}}a, Universidad de Antofagasta, Av.\ Angamos 601, Antofagasta, Chile  \label{M5}
\and
Centre for Electronic Imaging, Department of Physical Sciences, The Open University, Milton Keynes, MK7 6AA, UK  \label{M6}
\and
 Instituto de Astronom{\'{\i}}a y Ciencias Planetarias, Universidad de Atacama, Avenida Copayapu 485, Copiap\'o, Atacama, Chile  \label{M7}
\newpage
\and
Universit{\"a}t Hamburg, Faculty of Mathematics, Informatics and Natural Sciences, Department of Earth Sciences, Meteorological Institute, Bundesstrasse 55, 20146 Hamburg, Germany  \label{M8}
\and
Instituto de Astrofísica Pontificia Universidad Cat\'olica de Chile, Avenida Vicuna Mackenna 4860, Macul, Santiago, Chile  \label{M9}
\and
Department of Physics, Sharif University of Technology, PO Box 11155-9161 Tehran, Iran  \label{M10}
\and
Institute for Astronomy, University of Edinburgh, Royal Observatory, Edinburgh EH9 3HJ, UK  \label{M11}
\and
Astrophysics Group, Keele University, Staffordshire, ST5 5BG, ~~~UK  \label{M12}
\and
National Astronomical Observatories, Chinese Academy of Sciences, Beijing 100101, China  \label{L2}
\and
Las Cumbres Observatory, 6740 Cortona Drive, suite 102, Goleta, CA 93117, USA \label{L3}
\and
School of Physics and Astronomy, Tel-Aviv University, Tel-Aviv 6997801, Israel \label{L5}
\and
Auckland Observatory, Auckland, New Zealand\label{U1}
\and
Kumeu Observatory, Kumeu, New Zealand\label{U5}
\and
Universidade Federal do Rio Grande do Norte (UFRN), Departamento de F\'sica, 59078-970, Natal, RN, Brazil\label{U2}
\and
Laboratório Nacional de Astrofísica, Rua Estados Unidos 154,
37504-364, Itajubá - MG, Brazil\label{U9}
\and
Department of Physics, Chungbuk National University, Cheongju 28644, Republic of Korea\label{U10}
\and
Farm Cove Observatory, Centre for Backyard Astrophysics, Pakuranga, Auckland, New Zealand\label{U6}
\and
Klein Karoo Observatory, Centre for Backyard Astrophysics, Calitzdorp, South Africa\label{U7}
\and
Institute for Radio Astronomy and Space Research (IRASR), AUT University, Auckland, New Zealand\label{U8}
}

   \date{Received ; accepted }

 
  \abstract
   {Brown dwarfs are poorly understood transition objects between stars and planets, with several competing mechanisms having been proposed for their formation. Mass measurements are generally difficult for isolated objects but also for brown dwarfs orbiting low-mass stars, which are often too faint for spectroscopic follow-up.
}
   {Microlensing provides an alternative tool for the discovery and investigation of such faint systems. Here we present the analysis of the microlensing event OGLE-2019-BLG-0033/MOA-2019-BLG-035, which is due to a binary system composed of a brown dwarf orbiting a red dwarf.}
   {Thanks to extensive ground observations and the availability of space observations from {\it Spitzer}, it has been possible to obtain accurate estimates of all microlensing parameters, including parallax, source radius and orbital motion of the binary lens.}
   {After accurate modeling, we find that the lens is composed of a red dwarf with mass ${M_{1}=0.149\pm 0.010 M_{\odot}}$  and a brown dwarf with mass $\mathrm{M_{2}=0.0463\pm0.0031 M_{\odot}}$, at a projected separation of $a_\bot=0.585$ au. The system has a peculiar velocity that is typical of old metal-poor populations in the thick disk. Percent precision in the mass measurement of brown dwarfs has been achieved only in a few microlensing events up to now, but will likely become common with the {\it Roman} space telescope. }{}

   
   \titlerunning{Precision measurement of a brown dwarf mass by microlensing}
   \authorrunning{Herald et al.}

   \maketitle
%

\section{Introduction}

The low-mass end of main-sequence stars is conventionally set by the minimum mass needed to trigger hydrogen burning in the core. This corresponds to $0.078 M_\sun$ at solar metallicity \citep{Auddy2016}. Objects below this mass, classified as brown dwarfs (BDs), may still burn deuterium for a short phase and then cool down similarly to planets \citep{Burrows1997}. The distinction between planets and BDs is more vague because the deuterium-burning limit of $0.013_\sun$ \citep{Spiegel2011} has little impact on the global properties and evolution of these substellar objects. It is also known that the formation of BDs may occur by instability and collapse of gas clouds with the same Jeans mechanism that generates stars \citep{Bejar2001}. However, this mechanism rapidly becomes inefficient at low masses and may not fully account for the number of observed BDs \citep{Larson1992,Elmegreen1997}. On the other hand, bigger planets formed by core accretion may exceed the above-mentioned planet-BD threshold and be classified as BDs, although they are formed starting from a rock-ice core \citep{molli2012,Whitworth2007}. However, the lack of transiting BDs in Kepler discoveries and of BD companions to Sun-like stars hints at some migration or instability mechanism depleting planetary systems of overly massive objects \citep{Marcy2000,Grether2006,johns2011}. Following a different route, lower-mass stars may be naturally formed in binaries or multiples by turbulent fragmentation of the protostellar clouds, with some of the low-mass companions lying in the BD mass range \citep{padoan2004}. 

Confronting the efficiency of the alternative channels for the formation of BDs is difficult because these elusive objects are intrinsically dim and can only be detected when they are relatively young and hot \citep{close2007,lafren2007}. Space-borne instruments are well-suited to detect young BDs in the infrared \citep{spezzi2011,Mclean2003}, with a recent direct radio discovery of a BD having been reported by \citet{HKveda2020}. Similar complications exist for small companions to red dwarfs as well, as red dwarfs are typically too faint targets for spectroscopic follow-up \citep{sahlmann2011}. The mass and the nature of such companions thus remains poorly known. 

The low luminosity of brown and red dwarfs is not a limitation if observations rely on the light of some other sources, as in gravitational microlensing. Indeed, most of the lenses populating our Galaxy and causing magnification of background sources are believed to be low-mass stars or possibly BDs and even rogue planets \citep{Mroz2017,Mroz2019,Mroz2020,Kim2021,Ryu2021}. Therefore, microlensing provides a key to explore the low end of the mass function throughout our Galaxy \citep{Koshimoto2021}. However, the mass measurement of individual lenses is generally difficult because the information stored in a basic microlensing event is very degenerate. Subtle anomalies, higher order effects or additional observations are needed to resolve such degeneracies. Fortunately, this happens routinely for a sizeable fraction of microlensing events, leading to relatively good mass estimates of otherwise undetectable objects with a variety of methods \citep{An2002,Wyrzykowski2020}.

Concerning the discoveries of BDs with microlensing, the first isolated brown dwarf was recognized in the microlensing event OGLE-2007-BLG-224 \citep{Gould2009}. BDs as companions to stellar primaries were often discovered \citep{Bozza2012,Ranc2015}. Interestingly, several binary systems composed of BDs have been detected \citep{Choi2013}, as well as a case of a planet orbiting a BD just slightly below the hydrogen-burning limit \citep{Bennett2008}. The growing number of BD discoveries allows microlensing to confirm previous considerations about the existence of a BD desert at short periods around Sun-like stars, with some possible accumulation at intermediate periods \citep{Ranc2015}. Particularly appreciated are discoveries with very precise mass measurements: a better than 10\% accuracy has been reached several times \citep{Gould2009,Choi2013,Han2013,Ranc2015,Albrow2018}.

In this paper we report the analysis of OGLE-2019-BLG-0033/MOA-2019-BLG-035, a long timescale event well-covered by many ground telescopes, both surveys and follow-up, which allow a very good determination of parallax and finite-source effects. With the help of {\it Spitzer} data from space, the best model for this binary lens event including orbital motion is uniquely determined. With all this information, it is possible to achieve an exceptional precision in the mass measurement of the two components of the system, which turn out to be a red and brown dwarf at intermediate separation. In Section 2  we present the observations used for the analysis. In Section 3 we describe the modeling stages of the microlensing light curve. In Section 4 we deal with the source analysis using the color-magnitude diagram from the MOA telescope. In Section 5 we finalize the full interpretation of the model presenting the physical parameters. Section 6 contains a discussion on precise mass measurements by microlensing and their contribution to the understanding of BDs. We conclude with a brief summary in Section 7.

\section{Observations}

\begin{figure*}[t]         
\centering
\includegraphics[scale=0.7]{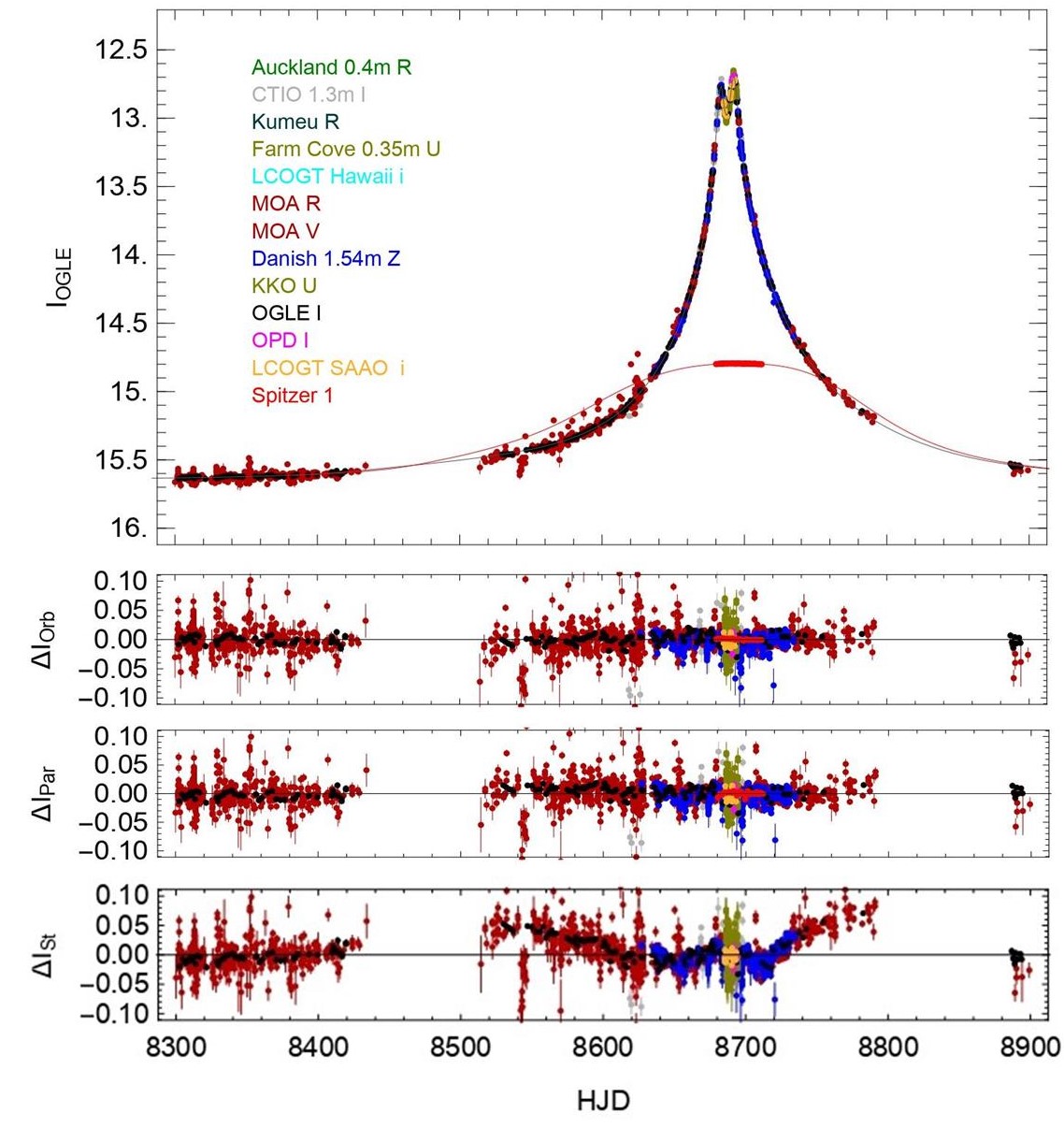}
\caption{Light curve of the event OGLE-2019-BLG-0033 showing all observations from different telescopes as described in the text. The black curve is the best microlensing model for ground observers, and the red curve is the best model for {\it Spitzer} observations as described in Section 3. In the bottom panel we show the residuals from several models: the best model including parallax and orbital motion, the best model with parallax without orbital motion, and the best static model without parallax.}\label{Fig lightcurve}%
\end{figure*}

The microlensing event OGLE-2019-BLG-0033/MOA-2019-BLG-035 was announced by the OGLE collaboration at the beginning of the 2019 bulge season on February 19 and independently found by MOA four days later. Its equatorial coordinates (J2000.0) are RA=$18:08:38.26$, Dec=$-30:03:38.7$, corresponding to Galactic coordinates $l=1.53^\circ$, $b=-4.90^\circ$. Fig. \ref{Fig lightcurve} shows all observations taken on OGLE-2019-BLG-0033 by different ground telescopes and from {\it Spitzer}. We summarize these observations and their reduction in this Section.

OGLE observations were carried out with the 1.3-m Warsaw telescope located at the Las Campanas Observatory, Chile. It was equipped with a 32-CCD mosaic camera covering 1.4 square degrees on the sky with a pixel scale of 0.26 arcsec/pixel. Most observations were obtained through the $I-$band filter (exposure time of 100 sec) for time series with occasional observations in the $V-$band for color information. The
lens is located in the BLG521 OGLE field which was observed with an average cadence of less than one observation per night. Photometry of OGLE-2019-BLG-0033 was derived using the standard OGLE photometric pipeline \citep{Udalski2015}, based on difference image analysis implementation by \citep{Wozniak2000}.

The MOA collaboration is carrying out a high cadence microlensing survey with a 1.8-m MOA-II telescope at Mt. John University Observatory in New Zealand \citep{Bond2001,Sumi2003}.  The telescope has a 2.2 deg$^2$ FOV, with a 10-chip CCD camera. The main observations are taken using the MOA-Red filter, which corresponds to the standard Cousins $R-$ and $I-$bands (630-1000 nm).
The MOA images were reduced with MOA’s implementation \citep{Bond2001} of the difference image analysis (DIA) method \citep{Tomaney1996,Alard1998,Alard2000}. In the MOA photometry, we de-trend the systematic errors that correlate with the seeing and airmass, as well as the motion, due to differential refraction, of a nearby, possibly unresolved star \citep{Bond2017}.

OGLE-2019-BLG-0033 was selected for {\it Spitzer} observations on 2019 May 10 at UT 04:11 ($HJD' = HJD-2400000 = 8613.67$), well before the binary anomaly was identified. It was selected as a "Subjective, Immediate" target with an "objective" cadence following the protocols described by \citet{Yee2015}. This cadence resulted in roughly one observation per day starting in Week 2 of the 2019 {\it Spitzer} campaign (the first week the target was observable). {\it Spitzer} observations were taken using the 3.6 $\mu$m ($L$-band) channel of the IRAC camera. Each observation consisted of six, dithered exposures. Because the target was very bright, the first ten epochs were taken with 12s exposures. Then, 30s exposures were used after HJD'=8692, once it was established that the target would not be saturated as seen from {\it Spitzer}. The {\it Spitzer} data were reduced using the photometry pipeline developed by \citet{Calchi2015} for IRAC data in crowded fields.

Because OGLE-2019-BLG-0033 was a {\it Spitzer} target outside of the KMTNet fields \citep{Kim2016}, and the OGLE cadence was $<1$ obs/night, follow-up telescopes have been particularly useful.

The Microlensing Follow-Up Network ($\mu$FUN) observed this event in the $I$- and $H$- bands using the SMARTS Cerro Tololo 1.3m telescope (CT13) in Chile. Initially, these observations were taken primarily in order to measure the $(I-H)$ color of the source. However, starting from HJD'=8666, this event was added to a group of events followed at a higher cadence in order to increase the sensitivity to small planets. Hence, OGLE-2019-BLG-0033 received a couple of observations per night from CT13 as time allowed. On 2019 July 19, $\mu$FUN recognized that the event was deviating from a point lens and sent out an anomaly alert at UT 17:09. As a result, additional follow-up observations were taken by several $\mu$FUN observatories, including Auckland Observatory (AO), Farm Cove Observatory (FCO), and Kumeu Observatory in New Zealand. AO and Kumeu observed using a Wratten $\# 12$ filter (designated $R$), while FCO observed without a filter (designated $U$ for ``unfiltered"). Observations were also obtained in $i$ band by the 1.6m telescope at Observatorio do Pico dos Dias (OPD) located in Brazil and using a clear filter (also designated $U$) from Klein Karoo Observatory in South Africa. All $\mu$FUN data, including CT13, were reduced using the DoPHOT pipeline \citep{Schechter1993}.

Las Cumbres Observatory (LCO) global network conducted observations using the 0.4m telescope at the South African Astronomical Observatory (SAAO) in South Africa and Haleakala Observatory in Hawaii (FTN). The LCO data were reduced using a custom DIA pipeline \citep{Zang2018} based on the ISIS package \citep{Alard1998,Alard2000}. 

The MiNDSTEp data were obtained using the Danish 1.54m Telescope at ESO La Silla Observatory in Chile, as part of the MiNDSTEp microlensing follow-up program \citep{Dominik2010}. The Danish 1.54m Telescope is equipped with a multi-band EMCCD instrument \citep{Skottfelt2015} providing shifted and co-added images in its custom red and visual passbands. This work utilizes red band time-series photometry which was reduced with a modified version of DANDIA \citep{Bramich2008,Bramich2013}. 817 observations were taken, of which three were discarded based on the photometric scale factors close to zero which serves as proxy to identify systematics in photometric measurements - most commonly overcast skies.

\section{Modeling} \label{Sec modeling}

Microlensing occurs when the lens passes close to the line of sight to a background source whose flux is thus temporarily magnified. The fundamental angular scale that governs microlensing is the angular Einstein radius
\begin{equation}
    \theta_E=\sqrt{\kappa M \pi_{\rm rel}}, \label{thetaE}
\end{equation}
where $M$ is the total lens mass, $\kappa=4G/(c^2\mathrm{au})$ combines Newton's constant, the speed of light and the astronomical unit, and 
\begin{equation}
    \pi_{\rm rel}=\mathrm{au}\left(\frac{1}{D_L}-\frac{1}{D_S}\right) \label{pirel}
\end{equation}
is the difference of the lens and source parallaxes.

For a lens moving with respect to the source with proper motion $\mu$, the timescale of the event (or Einstein time) is
\begin{equation}
    t_E=\frac{\theta_E}{\mu}.  \label{tE}
\end{equation}

The magnification of the source is maximum at the time of closest approach of the lens to the source line of sight $t_0$, when the minimum angular separation $u_0$ is reached (expressed in units of the Einstein radius).

If the lens is composed of two masses, besides $t_E$, $t_0$ and $u_0$, the light curve also depends on additional parameters: the mass ratio $q$, the projected angular separation in Einstein radii $s$, the position angle $\alpha$ of the second mass relative to the primary measured from the lens proper motion vector, and the source angular radius $\rho_*$ in units of the Einstein radius.

For long-timescale microlensing events, the motion of the Earth around the Sun must be taken into account, as it affects the apparent relative proper motion of lens and source. Such annual parallax effect is modeled by two additional parameters $\pi_{E,N}$ and $\pi_{E,E}$ representing the North and East components of the parallax vector, whose modulus is 
\begin{equation}
    \pi_E=\frac{\pi_{\rm rel}}{\theta_E} \label{piE}
\end{equation}
and whose direction is given by the lens-source proper motion direction \citep{Gould1992,Gould2000}. A binary lens model including parallax has thus 9 parameters.  

Finally, as the two lenses are gravitationally bound, they should orbit around their common center of mass. The full Keplerian motion can be modeled by 5 additional parameters \citep{Skowron2011}. Yet, because the duration of the microlensing event is often too short to derive a full orbit, it is generally sufficient to add a minimal set of 3 velocity components to obtain a circular orbit \citep{Skowron2011,Bozza2021}. A 2-component orbital motion, sometimes used for minimal fits, should be avoided as it leads to unphysical orbital trajectories \citep{Bozza2021,Ma2021}. The three components of the velocity are $\left(\gamma_1 \equiv \frac{ds}{dt}/s,\gamma_2=\frac{d \alpha}{dt},\gamma_3= \frac{ds_z}{dt}/s\right)$, where $s_z$ is the separation between the two lenses along the line of sight in Einstein radii. A binary lens model including parallax and orbital motion has thus 12 parameters.

As the lens moves in front of the source (or, equivalently, the source moves behind the lens), the source may enter regions in which new images are created. The boundaries of such regions are called caustics and determine the overall shape of the light curve for binary lenses.

\begin{figure}[th]         
\centering
\includegraphics[scale=0.75]{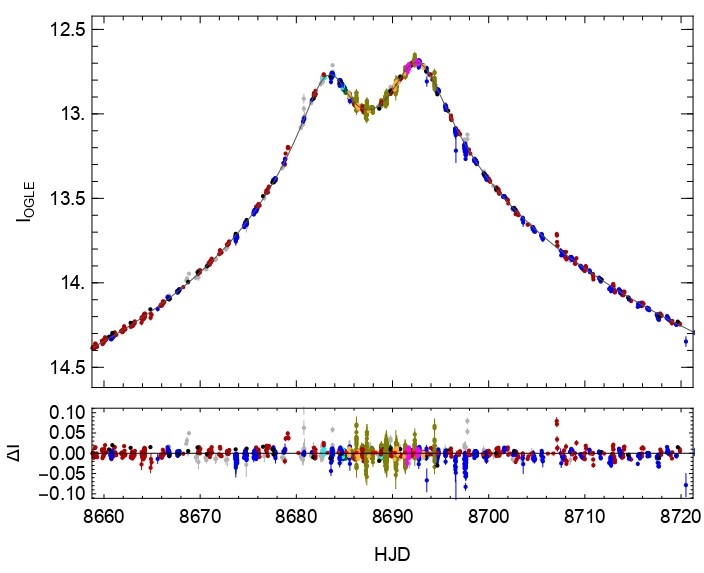}
\caption{Zoom on the double-peak region of the light curve. The color-coding for the observations is the same as in Fig. \ref{Fig lightcurve}.}\label{Fig zoom}
\end{figure}

\begin{figure*}[th]         
\centering
\includegraphics[width=14cm]{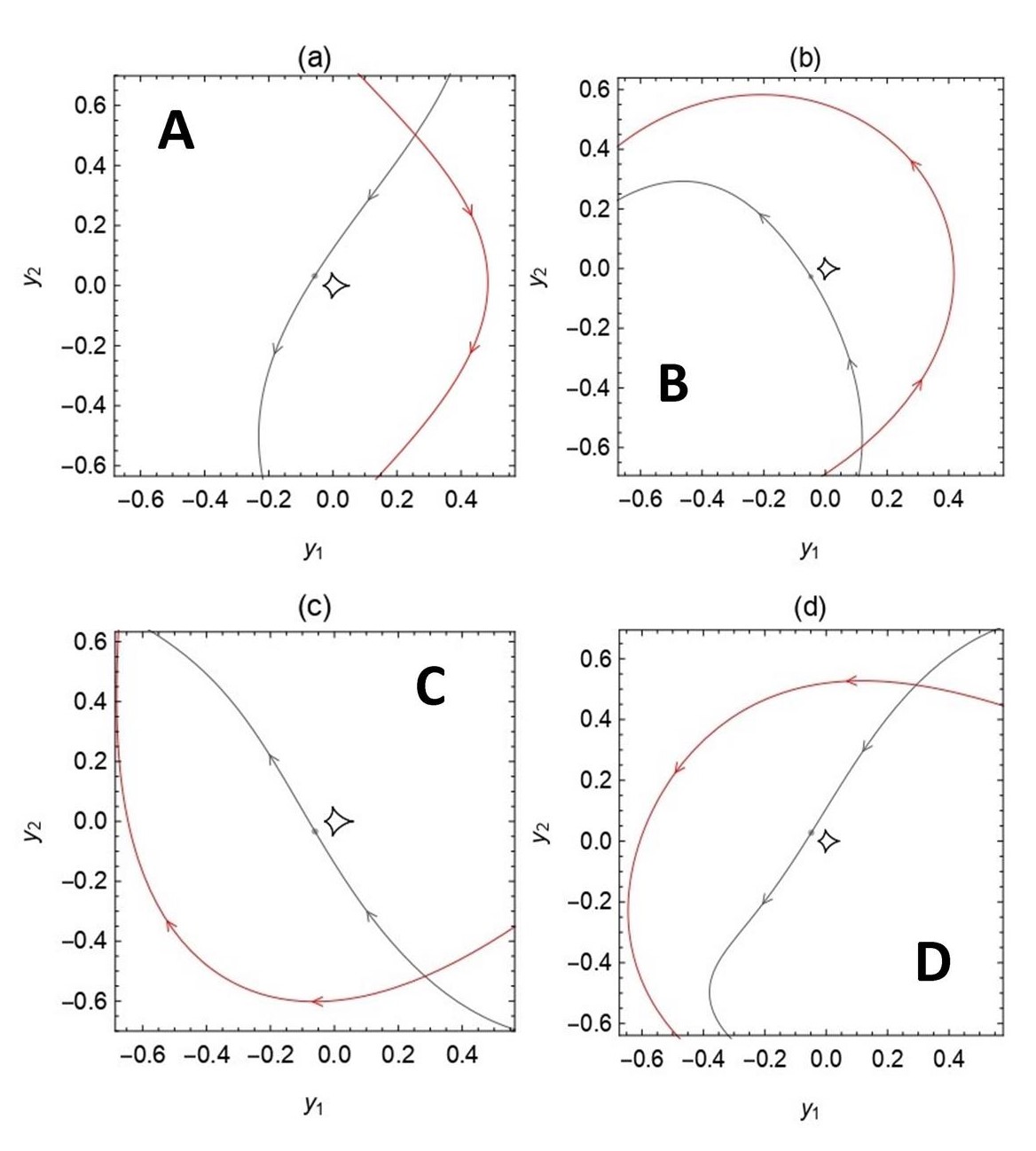}
\caption{Caustics of the four binary lens models examined in our analysis, with the best model labeled as A. The source trajectories are also shown as seen from Earth observatories (black) and from {\it Spitzer} (red).} \label{Fig caustic}
\end{figure*}

\begin{figure}[th]         
\centering
\includegraphics[width=7cm]{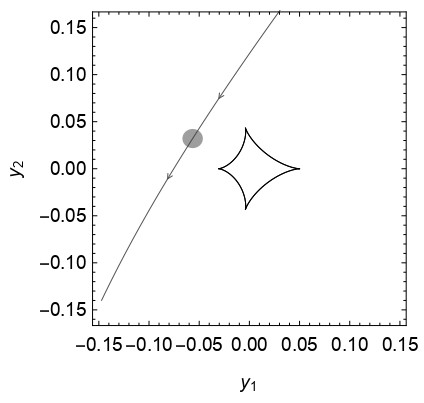}
\caption{A zoom-in of the caustic of the best binary lens model with the source trajectory and its size shown by the gray disk} \label{Fig causticzoom}
\end{figure}

Fig. \ref{Fig zoom} zooms on the double-peak region of the light curve of OGLE-2019-BLG-0033, occurring at moderate magnification ($A_{max}\simeq 15)$. This structure indicates that the central caustic generated by the lens has a typical astroid shape, which may arise either for close binary lenses or for a lens perturbed by a wide companion \citep{Dominik1999,Bozza2000}. Fig. \ref{Fig caustic} shows this shape as reconstructed by the best models to be described in this section, with a zoom-in in Fig. \ref{Fig causticzoom}.

%
The presence of a double-peak was immediately recognized during the observation campaigns, as noted in Section 2. Automatic modeling of the available online photometry by \texttt{RTModel}\footnote{\url{http://www.fisica.unisa.it/GravitationAstrophysics/RTModel.htm}. \texttt{RTModel} performs Levenberg-Marquardt fitting starting from initial conditions obtained by matching the data to template light curves from a library \citep{Mao1995}} found a full solution on 2019 August 25 including annual parallax and finite source effect, with the conclusion that the system was made of a red and brown dwarf, with masses very similar to those reported hereafter the full analysis.

\subsection{Detailed modeling procedure}

The photometry collected by all observatories has been reduced according to the procedures described in Section 2. We noted that the {\it Spitzer} light curve consists of 29 data points spanning 32 nights. These observations are around the peak of the magnification as seen from the ground observations and are very far from the baseline. Without a baseline, {\it Spitzer} observations for this event must be complemented by a flux constraint to be included in the analysis \citep{Calchi2015}. Given the special role of {\it Spitzer} data, which provide an independent determination of the parallax with respect to ground data, we decided to first analyze the ground data alone and obtain a first determination of all microlensing parameters of the event. As a second step, using the flux constraint on {\it Spitzer } data and comparing with the measured flux, we infer the geometry of the event as seen from the satellite alone. From this we obtain an independent estimate of the parallax that can be compared with the ground-only measurement to provide an important confirmation of the previous result \citep{Gould2012}. Finally, we will present results for a combined fit of ground + {\it Spitzer} data and discuss the impact of satellite data in the fit.

\subsubsection{Modeling of ground data} \label{Sec ground}

With all available ground data, we re-started an \texttt{RTModel} search and evaluated all possible competing models. This first run confirmed the preliminary model. However, it is well-known that parallax measurements using satellites are subject to a four-fold satellite degeneracy \citep{Refsdal1966,Gould1994b}, which corresponds to four competing models that can be obtained by reflection of the source trajectory around the binary-lens axis and by changes of signs in the parallax components. These four models have been named A, B, C and D as shown in Fig. \ref{Fig caustic}. We then decided to check all possibilities in parallel before dismissing any of them. Moreover, for all models we obtained a significant improvement by including orbital motion. It is important to correctly account for this last additional effect because it impacts the estimated components of the parallax \citep{Skowron2011,Batista2011}. We work in the geocentric frame setting the reference time for the parallax and orbital motion calculations as $t_{0,orb}=t_{0,par}=t_0$.

After this first step we have re-normalized the error bars of all datasets ensuring that $\chi^2/d.o.f.=1$ for the model with the lowest $\chi^2$. This standard procedure makes the fit more robust against possible low-level unknown systematics in the data \citep{Miyake2011}. However, we note that all datasets very accurately follow the best model with no particular deviations, as is evident from Figs. \ref{Fig lightcurve}-\ref{Fig zoom}. 

With the re-normalized uncertainties and including appropriate limb darkening coefficients for the source in each band, as detailed in Section 4, we have run Markov Chain Monte Carlo with 1 million samples to explore the parameter space around each model. As for \texttt{RTModel}, the microlensing magnification has been computed by the contour integration code \texttt{VBBinaryLensing}\footnote{\url{https://github.com/valboz/VBBinaryLensing}} \citep{Bozza2010,Bozza2018,Bozza2021}. The final parameters for each of the four models are listed in Table \ref{Tab models}, where we see that the model labeled $A$ stands out with a $\Delta \chi^2=120$ from the closest alternative, which is $B$. In this table we also include the baseline magnitude for OGLE ($I_{OGLE}$) and the relative blending fraction $BF_{OGLE}$, i.e. the ratio of the contaminating flux from unresolved stars in the blend to the source flux.

\begin{table*}[t]\begin{center}\centering
    \caption{Parameters of the best microlensing models found with ground-only data.}
       \label{Tab models}
       \renewcommand{\arraystretch}{1.5}
       \begin{tabular}{p{3cm}p{3cm}p{3cm}p{3cm}p{3cm}}
       \hline \hline
       \textbf{Parameter}          & A & B & C & D \\ \hline
       $\chi^2$ & $4793.2$ & $4913.0$ & $4919.6$ & $4921.6$ \\ \hline
       $s$ & $0.3325\pm{0.0024}$ & $0.3228_{-0.0056}^{+0.0008}$  & $0.3381_{-0.0035}^{+0.0016}$ & $0.3136_{-0.0042}^{+0.0012}$ \\
       $q$ & $0.3157\pm{0.0059}$ & $0.2712_{-0.0017}^{+0.0085}$ & $0.3360_{-0.0051}^{+0.0098}$ & $0.3076_{-0.0046}^{+0.0115}$ \\
       $u_0$ & $-0.06498\pm{0.00036}$ & $0.05479_{-0.00081}^{+0.00017}$ & $0.06899_{-0.00055}^{+0.00034}$ & $-0.05594_{-0.00029}^{+0.00044}$ \\
       $\alpha$ & $1.0311_{-0.0024}^{+0.0029}$ & $5.2539_{-0.0011}^{+0.0052}$ & $5.2480_{-0.0017}^{+0.0037}$ & $1.0173_{-0.0041}^{+0.0016}$ \\
       $\rho_{\star}$ & $0.01017\pm{0.0003}$ & $0.00857_{-0.00050}^{+0.00020}$ & $0.01083_{-0.00058}^{+0.00046}$ & $0.01042_{-0.00079}^{+0.00049}$ \\
       $t_E \hspace{2mm} (days)$ & $103.64\pm{0.57}$ & $119.26_{-0.25}^{+1.55}$ & $102.70_{-0.38}^{+0.71}$ & $121.56_{-0.60}^{+0.81}$ \\
       $t_0 \hspace{2mm} (HJD')$ & $8689.840_{-0.013}^{+0.017}$ & $8689.854_{-0.029}^{+0.005}$ & $8689.841_{-0.019}^{+0.011}$ & $8689.758_{-0.020}^{+0.008}$ \\
       $\pi_{E,N}$ & $0.2971_{-0.0082}^{+0.0033}$ & $-0.2566_{-0.0075}^{+0.0039}$ & $0.2742_{-0.0064}^{+0.0077}$ & $-0.2740_{-0.0033}^{+0.0119}$ \\
       $\pi_{E,E}$ & $-0.1962_{-0.0019}^{+0.0035}$ & $-0.1199_{-0.0010}^{+0.0042}$ & $-0.2021_{-0.0038}^{+0.0046}$ & $-0.1448_{-0.0018}^{+0.0045}$  \\
       $({ds}/{dt})/{s} \hspace{2mm} ({yr}^{-1}) $ & $-1.109_{-0.055}^{+0.079}$ & $-1.756_{-0.142}^{+0.051}$ & $-0.675_{-0.124}^{+0.051}$ & $-1.770_{-0.154}^{+0.050}$ \\
       $({d\alpha}/{dt}) \hspace{2mm} ({yr}^{-1}) $ & $-0.146_{-0.088}^{+0.084}$ & $1.729_{-0.097}^{+0.008}$ & $-2.369_{-0.149}^{+0.091}$ & $0.569_{-0.087}^{+0.179}$ \\
       $({ds_z}/{dt})/s \hspace{2mm} ({yr}^{-1}) $ & $<0.65$ & $<1.58$ & $<0.77$ & $<3.57$ \\ \hline
       $I_{OGLE}$ & $15.6463_{-0.0005}^{+0.0011}$ & $15.6397_{-0.0005}^{+0.0010}$  & $15.6505_{-0.0013}^{+0.0009}$ & $15.6409_{-0.0003}^{+0.0012}$ \\
       $BF_{OGLE}$ & $0.0461_{-0.0066}^{+0.0048}$ & $0.2529_{-0.0037}^{+0.0187}$  & $-0.0195_{-0.0046}^{+0.0081}$ & $0.2307_{-0.0077}^{+0.0094}$ \\ \hline
       
       \end{tabular}
\end{center}\end{table*}

For reference, the best model without orbital motion has $\Delta \chi^2=+477$ with respect to the best solution. The static model without parallax has $\Delta \chi^2=+4020$. The best binary source model with a single-lens has $\Delta \chi^2=+906$, while a model with a binary lens including parallax and xallarap (i.e., source moving around an unseen companion) gives $\Delta \chi^2=+215$.

The quality of the ground data, the coverage of the light curve, combined with the favorable case of a giant source with negligible blending and a long timescale ($t_E \simeq 103.6 d$), allow us to obtain particularly accurate estimates of all microlensing parameters. The event can be clearly ascribed to a close binary system with a secondary object $1/3$ as massive as the primary. The source size parameter $\rho_*$ is measured at $3\%$ precision, in spite of the fact that the source trajectory does not cross any caustics. This is due to the fact that the giant source passes over the magnified lobes surrounding two cusps of the astroid caustic. Its size is sufficient to be sensitive to the steep gradients in these regions, as shown in Fig. \ref{Fig causticzoom}. Both parallax components are also particularly accurate even using ground data alone without any continuous or discrete degeneracies \citep{Refsdal1966,Gould1994,Smith2003,Poindexter2005}. Finally, orbital motion is clearly measured in its first component $ds/dt$, it is marginally seen in $d\alpha/dt$, while only an upper limit can be given to the radial velocity component. The blending ratio is close to zero, with the flux vastly dominated by the magnified giant source, which makes the source analysis easier.

\subsubsection{Parallax determination from {\it Spitzer}}

In principle, parallax determination from the ground can be affected by competing higher order effects, such as orbital motion of the lens, xallarap, or even long-term source variability. So, an independent confirmation of the result using a different observation point is desirable. The existence of {\it Spitzer} data provides the opportunity to make such test and check the consistency of the results. In order to do this, we follow the cheap space-based parallax method suggested by \citet{Gould2012} and already tested by \citet{Shin2018,Shin2022}.

\begin{figure}[th]         
\centering
\includegraphics[width=9cm]{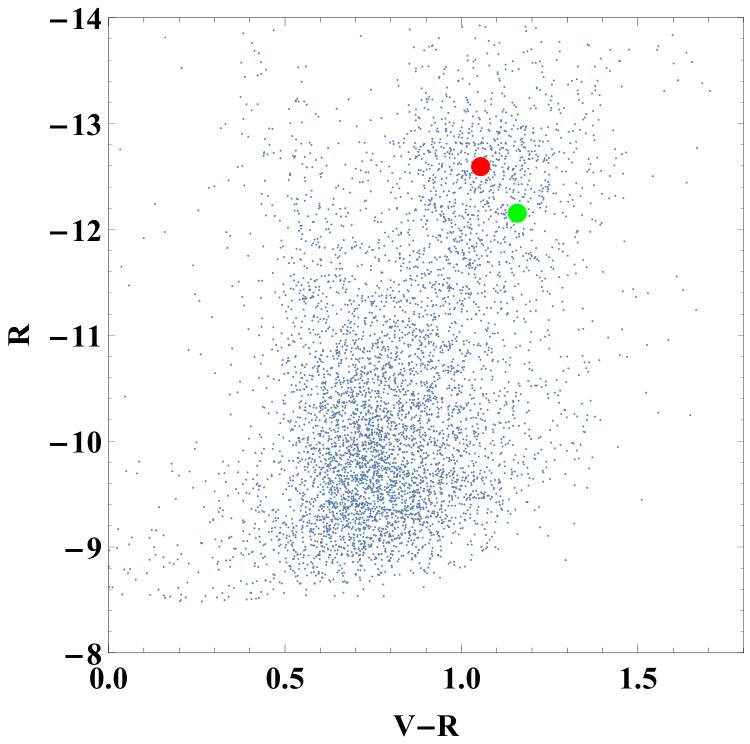}
\includegraphics[width=9cm]{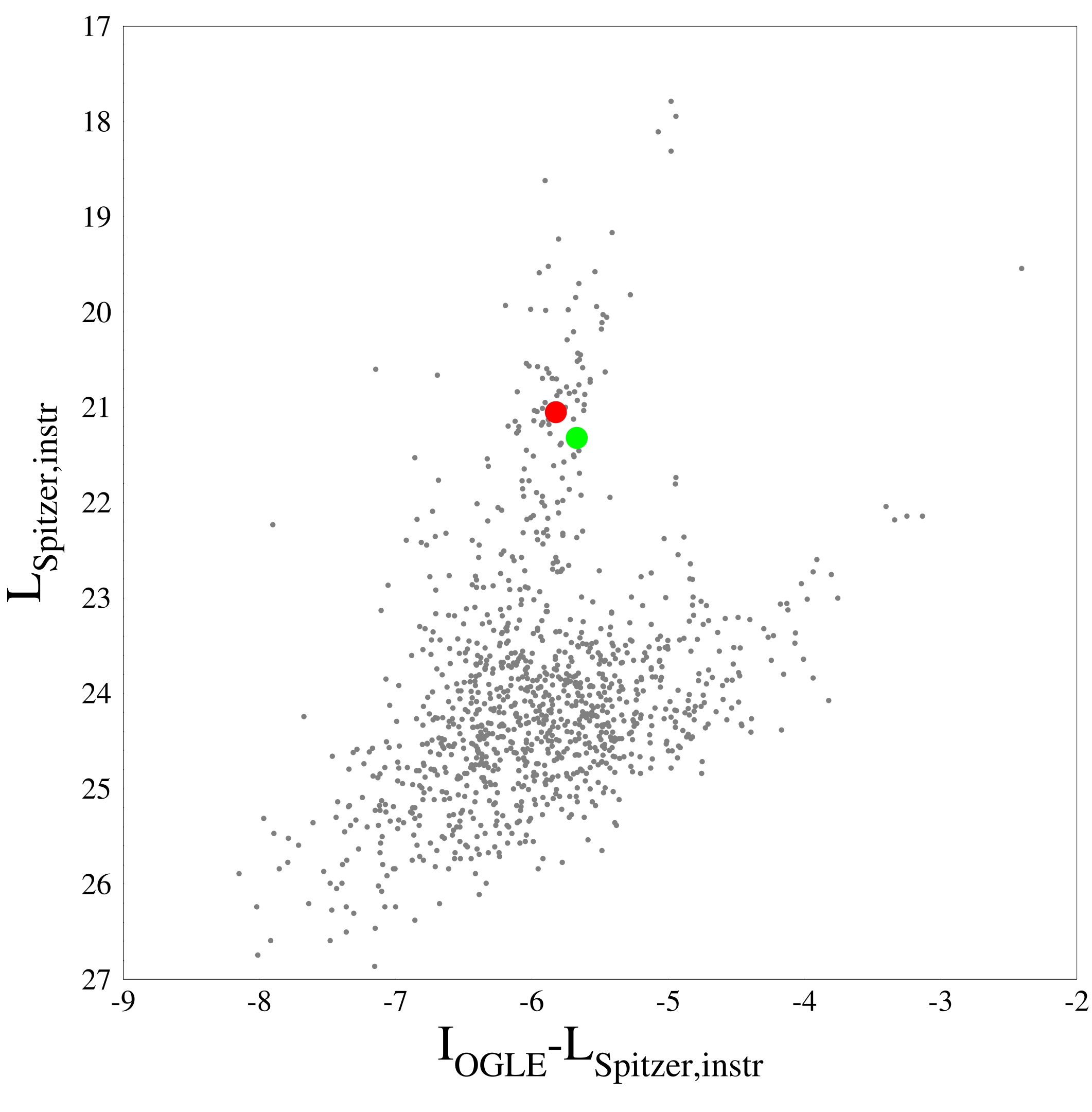}
\caption{Top panel: Color-Magnitude Diagram of stars in the $2^\prime$
 field of the event OGLE-2019-BLG-0033 built from MOA observations. The red dot corresponds to the center of the red clump and the green dot shows the position of the source. Bottom panel: CMD built from $I$-band observations from OGLE and $L$-band measurements from {\it Spitzer}.}\label{Fig CMD}
\end{figure}

The ground model confirms that the blending in OGLE photometry, if any, is negligible. {\it Spitzer}, however, has a pixel scale of 1.2$''$ compared to 0.26$''$ for OGLE, being more exposed to blending by nearby objects. Fortunately, no stars within this angular distance appear in OGLE images or in OGLE catalog and indeed the source appears well isolated in {\it Spitzer} images. As a further proof that the source does not suffer from contamination in {\it Spitzer images}, we compare the Color-Magnitude-Diagram (CMD) obtained using MOA observations in $V$ and $R$ bands (top panel of Fig. \ref{Fig CMD}) with a CMD obtained using OGLE $I$-band and {\it Spitzer} $L$-band (bottom panel of Fig. \ref{Fig CMD}). In both cases, the source lies just slightly below the centroid of the red clump, demonstrating that the ground and space measurements refer to the same object with no appreciable blending. 

Following the strategy outlined by \citet{Yee2013} and based upon {\it Spitzer} photometry of field stars cross-matched with OGLE-EWS CMD, we evaluate a corresponding color $I-L=-5.67 \pm 0.06$ for a Zero Point at 25 for {\it Spitzer}. With an OGLE baseline of $I=15.65$, this translates to a baseline flux for {\it Spitzer} 
\begin{equation}
f_{base,{\it Sp}}=29.65\pm 0.82 \label{fSpitzer}
\end{equation}
in instrumental units.

On the other hand, the {\it Spitzer} measurements during the observation window show a quite flat light curve. In particular, at time $t_0$, the two closest observations average at $f_{t_0,{\it Sp}}=69.52 \pm 0.26$. This means that the magnification as seen from {\it Spitzer} at this time is
\begin{equation}
    A_{0,{\it Sp}}=\frac{f_{t_0,{\it Sp}}}{f_{base,{\it Sp}}}=2.34 \pm 0.07,
\end{equation}
where again we are assuming that blending is negligible, as indicated by the ground-data model.

If the lens were a point mass, we would simply invert the Paczynski formula 
\begin{equation}
A(u)=\frac{u^2+2}{u\sqrt{u^2+4}}
\end{equation}
to obtain the angular separation of source and lens in Einstein radii
\begin{equation}
    u(A)= \sqrt{2\left[(1-A^{-2})^{-1/2} - 1\right]}.
\end{equation}

The so-derived separation is $|u_{0,\it Sp}|=0.460 \pm 0.015$. Since our lens is binary, we can make a more detailed search for angular separations yielding a magnification equal to $A_{0,{\it Sp}}$. We find that $0.44<|u_{0,\it Sp}|<0.48$ depending on the orientation of the source as seen from {\it Spitzer} with respect to the binary lens axis. This is very similar to the result for a single lens quoted before. Indeed, {\it Spitzer}'s separation is 7 times larger than the size of the caustic in Einstein units, so that perturbations of the magnification from lens binarity are small.

The offset of the source as seen from {\it Spitzer} with respect to the ground again depends on the unknown relative position angle. Therefore, this offset may range from $|u_{0,\it Sp}|-|u_0|$ to $|u_{0,\it Sp}|+|u_0|$. Combining the uncertainties, we set $\Delta u_0=|u_{0,\it Sp}-u_0|=0.46 \pm 0.07$.

At time $t_0$, the separation of {\it Spitzer} from Earth projected orthogonally to the line of sight was $d_{0,{\it Sp}}=1.51$ au. So, we finally obtain
\begin{equation}
    \pi_{E,Sp}=\Delta u_0\frac{\mathrm{au}}{d_{0,{\it Sp}}} =0.304 \pm 0.05.
\end{equation}

This can be compared to the result from the ground-only fit, which corresponds to
\begin{equation}
    \pi_{E,gr}=0.356 \pm 0.006.
\end{equation}

We stress that the only information from the ground model used in the derivation of the {\it Spitzer} parallax was the time of the peak $t_0$, the source separation from the lens $u_0$ and the absence of blending, so, for what concerns the parallax, the two results can be considered completely independent of each other. The consistency becomes even more evident when we note that for the best model $A$ the predicted offset of the source $\delta u$ as seen from {\it Spitzer} is indeed close the maximal value quoted before. We conclude that {\it Spitzer} fully validates the ground-only derived parallax making our result more robust against any possible sources of systematics that would remain uncontrolled with ground-only data.

\subsubsection{Models including ground and {\it Spitzer} data}

\begin{figure}[th]         
\centering
\includegraphics[width=9cm]{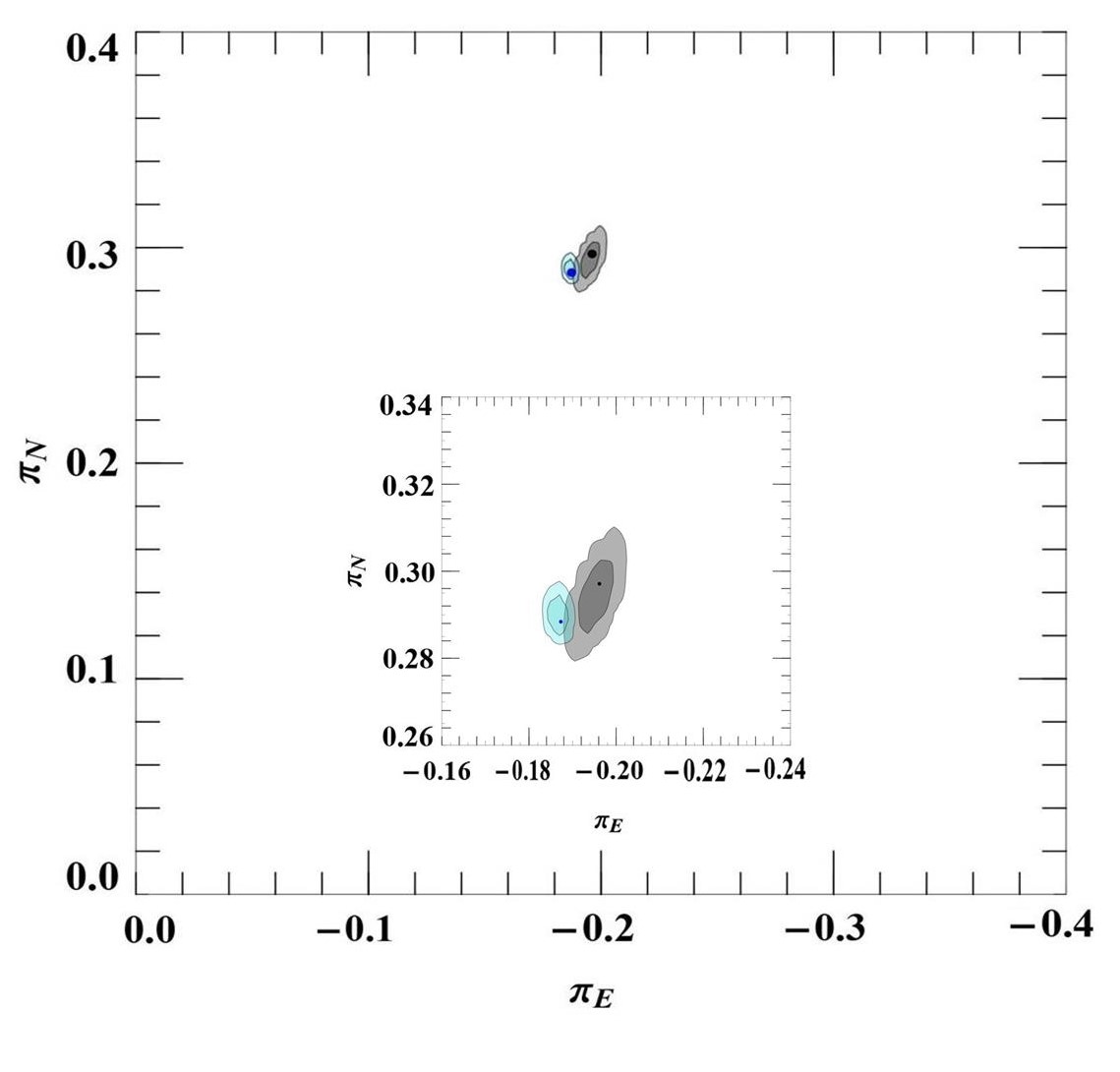}
\caption{Components of the parallax vector as found by the fit excluding {\it Spitzer} (in gray) or including {\it Spitzer} data \label{Fig parallax} (in cyan). Confidence levels at $68\%$ and $95\%$ are given.}
\end{figure}

As a final check for the consistency between ground and {\it Spitzer} observations, we conducted Markov chain explorations of the parameter space including {\it Spitzer} data for all four configurations studied in Section \ref{Sec ground}. In these fits we have forced {\it Spitzer} flux to fulfill the constraint (\ref{fSpitzer}) so as to keep the exploration within physically acceptable regions. The results are summarized in Table \ref{Tab Spitzer}, where we see that model $A$ is largely preferred also by {\it Spitzer} data, with a $\Delta \chi^2=46$, compared to $\Delta \chi^2>71$ for other models. The value of the parallax parameter $\pi_E$ remains within $2\sigma$ for model $A$, while it is significantly altered in other models, proving that the space parallax would be in tension with the ground parallax in these cases. This tension is apparent in the light curves of these models, which predict a declining trend for the {\it Spitzer} light curve that is not observed. The {\it Spitzer} light curve is quite flat in the observation window, as correctly predicted by model $A$ as shown in Fig. \ref{Fig lightcurve}. Therefore, {\it Spitzer} provides an additional strong confirmation of the model found using ground data only. In Fig. \ref{Fig parallax} we can appreciate how accurate the parallax measurement is for our event and the consistency of the fits performed with or without {\it Spitzer} data in model $A$.

\begin{table}[th]\begin{center}
    \caption{Comparison of parallax $\pi_E$ and $\chi^2$ for our four models if {\it Spitzer} data are included or excluded.}
       \label{Tab Spitzer}
\renewcommand{\arraystretch}{1.7}
\begin{tabular}{p{3.5cm}p{2.3cm}p{1.2cm}}
\hline \hline
\textbf{Model}          &  \textbf {$\pi_E$} &  \textbf {$\chi^2$}       \\ \hline 
\multicolumn{1}{l}{{\it Spitzer} Data Excluded}\\
$A$  &  $0.3560 \pm{0.0061}$ & 4793.2 \\
$B$  & $0.2832_{-0.0040}^{+0.0060}$&4913.0  \\
$C$  &  $0.3406_{-0.0068}^{+0.0078}$&4919.6 \\
$D$   & $0.3099_{-0.0121}^ {+0.0032}$ & 4921.6  \\\hline
\multicolumn{1}{l}{{\it Spitzer} Data Included}\\
$A$  &  $0.3439\pm{0.0005}$ & 4839.05 \\
$B$  & $0.3333_{-0.0008}^{+0.0005}$ & 5001.0  \\
$C$  & $0.2878_{-0.0011}^{+0.0008}$ & 5231.0\\
$D$   & $0.2839_{-0.0005}^ {+0.0011}$ & 4992.0   \\ \hline
\end{tabular}
\end{center}\end{table}

Taken the other way round, the full consistency of ground and {\it Spitzer} parallaxes also demonstrates that in this case {\it Spitzer} photometry was free of any important systematic effects, which may be present when the source is faint or blended \citep{Koshimoto2020} and that a correct use of the color constraint makes {\it Spitzer} data extremely useful to validate ground data and exclude possible additional effects.

In the following sections, we have the choice to use the parameters derived from ground-only fits or from ground-{\it Spitzer} combined fits. They give practically interchangeable results, with a slightly smaller uncertainty if we include {\it Spitzer} data. Therefore, we adopt the values of the combined fits given by Table \ref{Tab OBP-Spitzer} as reference for our analysis.

\begin{table}[th]  \begin{center}\centering
    \caption{Microlensing parameters for model A including {\it Spitzer} data}
       \label{Tab OBP-Spitzer}
       \renewcommand{\arraystretch}{1.6}
       \begin{tabular}{p{4cm}p{4cm}}
       \hline \hline
       \textbf{Parameter}    & \textbf{Model A w/ {\it Spitzer}} \\ \hline
       $\chi^2$ & $4839.05$  \\ \hline
       $s$ & $0.3336\pm{0.0024}$  \\
       $q$ & $0.3114\pm{0.0059}$ \\
       $u_0$ & $-0.06491\pm{0.00035}$  \\
       $\alpha$ & $1.0338_{-0.0023}^{+0.0030}$ \\
       $\rho_{\star}$ & $0.01007\pm{0.00035}$  \\
       $t_E \hspace{2mm} (days)$ & $103.85\pm{0.47}$ \\
       $t_0 \hspace{2mm} (HJD')$ & $8689.856_{-0.011}^{+0.015}$  \\
       $\pi_{E,N}$ & $0.2884_{-0.0006}^{+0.0010}$  \\
       $\pi_{E,E}$ & $-0.1873_{-0.0009}^{+0.0018}$  \\
       $({ds}/{dt})/{s} \hspace{2mm} ({yr}^{-1})$ & $-1.080_{-0.055}^{+0.084}$  \\
       $({d\alpha}/{dt}) \hspace{2mm} ({yr}^{-1})$ & $-0.091_{-0.073}^{+0.091}$  \\
       $({ds_z}/{dt})/s \hspace{2mm} ({yr}^{-1})$ & $<0.45$   \\ \hline
       $I_{OGLE}$ & $15.6459_{-0.0005}^{+0.0006}$  \\
       $BF_{OGLE}$ & $0.0477_{-0.0071}^{+0.0040}$  \\ \hline
       
       \end{tabular}
\end{center} 
\end{table}

\section{Source analysis}

The source characterization is important to determine the correct limb darkening profile to be used in the modeling of the light curve and the angular source radius $\theta_*$, which provides a physical scale hooked to the Einstein radius. In the case of OGLE-2019-BLG-0033, we have no observations from OGLE in $V$ band. Yet, we can exploit MOA observations in $V$ and $R$ bands to construct the color-magnitude-diagram shown in the top panel of Fig. \ref{Fig CMD}, in which we can place the source (green) and identify the center of the red clump on the giant branch (red). We have  $R_{MOA,Clump}=-12.5931 \pm{0.0091}$ and $(V-R)_{MOA,Clump}=1.0543 \pm{0.0065}$, which can be converted to standard Johnson-Cousins magnitudes using the photometric relations by \citet{Bond2017}:
\begin{eqnarray}
&I_{Clump}=&R_{MOA,Clump}+28.0264 \nonumber\\&&-0.1984*(V_{MOA,Clump}-R_{MOA,Clump})
\end{eqnarray}
\begin{eqnarray}
&V_{Clump}=&V_{MOA,Clump}+28.6274 \nonumber\\&&-0.1682*(V_{MOA,Clump}-R_{MOA,Clump}).
\end{eqnarray}

Hence, we obtain $I_{Clump}=15.2241 \pm{0.0095}$ and $(V-I)_{Clump}=1.6872 \pm{0.0074}$. Comparing to 
the Red Clump intrinsic magnitude ${I_{Clump,O}}=14.384  \pm{0.040}$ \citep{Nataf2013} and color ${(V-I)_{Clump,O}}=1.06 \pm{0.07}$ \citep{Bensby2011} at the Galactic coordinate of our microlensing event, we find a reddening of $E (V-I)=0.627$ and an extinction $A_I=0.852$. 

The best microlensing model indicates that the blending fraction is compatible with zero, so we attribute the baseline flux entirely to the source, as shown in Fig. \ref{Fig CMD}. Applying the same transformations to the source flux and taking into account the extinction and reddening just derived, we obtain ${(V-I)_{*,0}}=1.137 \pm{0.071}$ and ${I_{*,0}}=14.835 \pm{0.042} $. Following \citet{yoo2004}, we transform the $(V,I)$ bands to $(V,K)$ bands using the relations by \citet{Bessel1988} and then find the angular radius of the source following \citet{Kervella2004}
\begin{equation}
\theta_{\star} = 5.49 \pm{0.32} \; \mu as.
\end{equation}

The parallax from {\it Gaia } EDR3\footnote{\url{https://gea.esac.esa.int/archive/}} for our source is $\pi_S=-0.013\pm 0.089$ mas, thus compatible with zero within the errors \citep{Gaia2016,Gaia2021}. So, for an estimate of the source distance, we solely rely on the CMD. As the source position in the CMD is very close to the bulge red clump, it is reasonable to assume it is a bulge giant. Therefore, adopting the Galactic model by \citet{Dominik2006}, for the Galactic coordinates of OGLE-2019-BLG-0033 we find that the peak stellar density in the bulge along the observation cone is encountered at a distance $D_S=8.1$ kpc, which we assume to be a valid proxy for the source distance as well. The uncertainty in the source distance is assumed to be 1 kpc, reflecting the FWHM of the stellar density distribution along the line of sight.

In order to estimate the limb darkening coefficients for our source, we simulate a stellar population with IAC-Star \citep{Aparicio2004} with the stellar evolution library by \citet{Bertelli1994} and the bolometric correction by \citet{Castelli2003}. The best match with our source magnitude and color is found for $T_{eff}=4950 \hspace{0.5mm} K$ , $\log g= 2.77$ and $Z=0.011$. Using the tables by \citet{Claret2011}, we get the linear limb darkening coefficients in the relevant bands: $u_I= 0.5015$, $u_R=0.5983$, $u_V= 0.6945$. These coefficients have been used to obtain the microlensing models presented in Section \ref{Sec modeling}.

\section{Lens System Properties}

\subsection{Mass and distance}

For OGLE-2019-BLG-0033 we have an optimal circumstance in which the lens model is singled out without any degeneracies and with very accurate values of parallax and source size parameters. In addition, the source is a red clump giant with negligible blending flux, which allows an easy derivation of the angular source radius, useful to fix the Einstein radius as
\begin{equation}
    \theta_E=\frac{\theta_{\star}}{\rho_{\star}}=0.545 \pm{0.037} \; \mathrm{mas}. \label{thetaEnum}
\end{equation}

By inversion of Eqs. (\ref{thetaE}) and (\ref{piE}), we can calculate the total mass and the distance to the lens:
\begin{equation}
M=\frac{\theta_E}{\kappa\pi_E} = 0.195\pm{0.013} \hspace{1mm} M_{\odot}
\end{equation}
\begin{equation}
    D_L=\frac{\mathrm{au}}{\theta_E\pi_E + \pi_S}=3.22\pm 0.21 \; \mathrm{kpc} \label{DL}
\end{equation}
where $\pi_S=\mathrm{au}/D_S$ is the source parallax.

The masses of the two components of the binary lens can be found by use of the mass ratio $q$, which is very precisely fixed by the microlensing model: $M_1={M}/{(1+q)}$ and $M_2={qM}/{(1+q)}$. Finally, the projected separation of the two lenses can be obtained as
\begin{equation}
    a_\perp= s \theta_E D_L.
\end{equation}
 
The results of our analysis are reported in Table \ref{tab final}, showing that the binary system is composed of a red dwarf of $0.14 M_\sun$ and a BD of $0.046 M_\sun$ at a projected separation of $0.58$ au. The light from the system is very weak compared to the background microlensed source ($V\sim 26$ for a M5V red dwarf, as follows from \citet{Benedict2016}) in agreement with the negligible blending flux found in the model. The separation of half-au is quite typical for binary systems discovered through the microlensing method as the sensitivity to companions is maximized for separations of the same order as the Einstein radius. 

\begin{table}[th]\begin{center}
    \caption{Parameters of the binary lens system.}
       \label{tab final}
\renewcommand{\arraystretch}{1.7}
\begin{tabular}{ll}
\hline \hline
\textbf{Parameter}          &  \textbf {Value}        \\ \hline 
$M_1 \hspace{1mm} (M_{\odot})$  &  $0.1494 \pm{0.0099}$  \\
$M_2 \hspace{1mm} (M_{\odot})$  & $0.0463 \pm{0.0031}$   \\
$a_{\perp} \hspace{1mm} (au)$    &   $0.585\pm{0.054} $\\ 
$D_L \hspace{1mm} (kpc)$        &   $3.22\pm{0.21}$ \\
\hline
\end{tabular}
\end{center}\end{table}

\subsection{Orbital motion}

We have seen that orbital motion has been detected for our lens at least in the component along the binary lens axis. In microlensing events, a change in the separation $s$ is reflected in rapid evolution of the caustics, which leave an immediate imprint on the light curve. Therefore, it is expected that the component $\gamma_1=(ds/dt)/s$ is best constrained. The rotation of the axis $\gamma_2 = d\alpha/dt$ is compatible with zero at $1\sigma$ level, while for the radial component of the velocity we only have an upper limit, as typical in most microlensing events. With this scarce information, we may still check that the system is really bound by comparing the projected kinetic energy to the potential energy, namely a bound system must have
\begin{equation}
    K= \frac{(\gamma_{1}^2+\gamma_{2}^2)s^3\theta_E^3 D_{L}^3}{2GM}<1
\end{equation}
Using the values for our lens system, we find $K=0.0153$, which satisfies the constraint but is even relatively smaller than typical expectations from a random distribution of orbits. Such small values indicate a nearly edge-on orbit, which would apply to our case, given that $\gamma_2/\gamma_1=0.09$. So, with the information in hand, we can conclude that the orbital motion suggested by the light curve fit is perfectly acceptable and consistent with the constraints on the mass and scale of the system coming from the combination of parallax and finite source effects.

\subsection{Lens kinematics}

With the determination of the Einstein radius $\theta_E$ (\ref{thetaEnum}), we can find the relative lens-source proper motion from Eq. (\ref{tE}) as
\begin{equation}
    \mu_{\rm rel}=\frac{ \theta_E}{t_E}=1.92\pm{0.13} \; \mathrm{mas} \; \mathrm{yr}^{-1},
\end{equation}
which is relatively slow for a lens in the disk \citep{Han2003}. The components in the East and North directions in the geocentric frame can be derived from the parallax vector
\begin{equation}
   \mathbf{\mu}_{\rm rel,geo}= \frac{\mu_{\rm rel}}{\pi_E}(\pi_{E,E},\pi_{E,N}) = (-1.04 \pm 0.07, 1.61 \pm 0.11) \; \; \mathrm{mas/yr}.
\end{equation}

These can be easily transformed to the heliocentric frame using the velocity components of the Earth at time $t_0$ projected orthogonally to the line of sight
\begin{equation}
   \mathbf{\mu}_{\rm rel,hel}= \mathbf{\mu}_{\rm rel,geo} +\mathbf{v}_\oplus \frac{\pi_{\rm rel}}{\mathrm{au}} = (-0.04 \pm 0.07, 1.56 \pm 0.11) \; \; \mathrm{mas/yr}.
\end{equation}

Thanks to the {\it Gaia} EDR3 measurement of the proper motion of the source, we are in the position to make a full investigation of the lens kinematics and assign the lens to a distinct component of the Milky Way \citep{Gaia2016,Gaia2021}. From {\it Gaia}, we have
\begin{equation}
    {\mathbf \mu}_S = (-0.872 \pm 0.093, -7.28 \pm 0.067) \; \; \mathrm{mas/yr},
\end{equation}
with components given in the East and North directions respectively. Then, we can extract the lens proper motion
\begin{equation}
    \mathbf{\mu}_L=\mathbf{\mu}_{\rm rel,hel}+\mathbf{\mu}_S = (-0.91 \pm 0.12, -5.72 \pm 0.13) \; \; \mathrm{mas/yr} .
\end{equation}

We then rotate this vector by $61.36^\circ$ so as to measure its components in a Galactic frame
\begin{equation}
    \mathbf{\mu}_{L,gal}= (-5.46 \pm 0.12, -1.94 \pm 0.12) \; \; \mathrm{mas/yr} .
\end{equation}
Here, the first component is along the Galactic longitude direction $l$ and the second component is along the Galactic latitude $b$. As we know the distance of the lens (Eq. \ref{DL}), we can translate the proper motion to the velocity components
\begin{equation}
    \mathbf{v}_{L,gal}= (-83.2 \pm 5.7, -29.6 \pm 2.7) \; \; \mathrm{km/s}.
\end{equation}
Finally, subtracting the peculiar velocity of the Sun, we may move to the Local Standard of Rest (LSR)
\begin{equation}
    \mathbf{v}_{L,LSR}= (-71.0 \pm 5.7, -23.3 \pm 2.7) \; \; \mathrm{km/s}.
\end{equation}

Since the line of sight is very close to the Galactic center, these components are very close to the peculiar velocity components of the lens along the tangential circle $v$ and orthogonal to the Galactic plane $w$ respectively. A value of $v\sim -71$ km/s is quite typical of red metal-poor old stellar populations from the thick disk, as can be inferred from studies of the asymmetric drift \citep{Golubov2013}. So, the kinematic study allows us to firmly assign our lens, made up of a red and a brown dwarf, to Pop. II stars in the thick disk. Similar conclusions were obtained by \citet{Gould1992}, proving the effectiveness of microlensing in the investigation of populations of very low-mass components of our Galaxy.

\begin{table*}[th]
 \centering
  \caption{Binary microlensing events with relative error less than $10\%$ for the BD mass. The suffix A, B in the names indicates that the BD is the primary or the secondary component in the lens. No suffix means that the BD lens was isolated.}
    \begin{tabular}{lccl}
    \hline 
 Name Of The Event & {BD mass ($M_\sun$)} & Relative uncertainty ($\%$)  & Reference \\ \hline
OGLE-2011-BLG-0420A & $0.025$ & $4$ & \citep{Choi2013} \\ MOA-2007-BLG-197B &   $0.039$ & $5$ & \citep{Ranc2015} \\ OGLE-2009-BLG-151A & $0.018$ & $5.5$ & \citep{Choi2013} \\ \textbf{OGLE-2019-BLG-0033B} & $\mathbf{0.046}$ & $\mathbf{6.8}$ & \textbf{this work} \\
OGLE-2007-BLG-224 & $0.056$ & $7.1$ & \citep{Gould2009} \\ OGLE-2012-BLG-0358A & $0.022$ & $8.6$ & \citep{Han2013}  \\
OGLE-2016-BLG-1266A & $0.015$  & $10$ &{\citep{Albrow2018} } \\ 
MOA-2011-BLG-149B & $0.019$ & $10.5$ & \citep{Shin2012} \\ 
\hline
    
    \end{tabular}
  \label{tab microlensing BDs}
 
\end{table*}

\section{Discussion}

\subsection{Precision mass measurements by microlensing}

There is a good number of systems similar to OGLE-2019-BLG-0033 discovered in binary microlensing events, showing that such low-mass binary systems are very common in the Milky Way \citep{Ranc2015}. It is interesting to compare the precision of the mass measurement for our BD to that achieved in other similar microlensing events. Table \ref{tab microlensing BDs} collects the masses and the relative uncertainties of some microlensing events with binary lenses containing a BD. These events have been selected by us as featuring an uncertainty lower than $10\%$ in the BD mass. We then realize that our measurement is well-ranked as one of the most precise ever realized for a BD in a binary system. 

OGLE-2019-BLG-0033 represents a nice example showing the potential of microlensing observations to contribute to a detailed knowledge of the properties and statistics of low-mass objects in our Galaxy, from planets to BDs and red dwarfs, but also stellar remnants \citep{Blackman2021,Wyrzykowski2020}. Nevertheless, obtaining a unique lens model without degenerate alternatives and with very precise values for the parameters is not that simple. The case of OGLE-2019-BLG-0033 shows that long events with clear parallax and orbital motion signals are optimal for at least two reasons: a precise parallax detection gives a mass-distance relation to be combined with other constraints on $\theta_E$; orbital motion may distinguish otherwise degenerate solutions and help single-out the correct model. Short events, instead, are typically affected by discrete degeneracies that leave several alternative interpretations for the lens geometry with typically different values for the masses in spite of individual low uncertainties for the degenerate models \citep{Shvartzvald2016,Mroz2020}. However, annual parallax measurements rely on long-term modulations in the observed flux for which there might be possible alternative explanations or contaminants, including lens orbital motion itsef, xallarap, long-term variability of the source, systematics in the data. Therefore, the presence of measurements from a different point of observation such as {\it Spitzer}, allows an independent determination of parallax that goes back to pure geometry rather than subtle modulations in the photometry. And in fact {\it Spitzer} observations contribute to further reduce the uncertainty of our best model.

The second ingredient to obtain a precise mass measurement is a good estimate of the Einstein angle $\theta_E$. In the case of OGLE-2019-BLG-0033 this is obtained by the detection of finite source effects in the light curve. Although the source did not cross any caustics, it was big enough that even a close approach with the cusps was sufficient to obtain a precise value of $\rho_*$. In addition, the source was a red clump giant with no blending, which made the source analysis particularly easy and precise. The importance of the source analysis should not be underestimated in microlensing mass measurements. Indeed, even in our optimal situation, $\theta_*$ dominates the error budget in the derived masses. One way to improve the source knowledge could be a systematic spectroscopic survey of bright sources of microlensing events, which may definitely enhance the significance of microlensing mass measurements \citep{Bensby2010,Bensby2011,Bensby2021}. An alternative to finite-source effects, precise mass-distance relations, can be obtained by high-resolution imaging, which works in a complementary way to annual parallax, as it privileges fast-moving lenses \citep{Bhattacharya2018}, but requires sufficiently bright lenses. Otherwise, only upper limits can be obtained. Measurements of $\theta_E$ have been recently obtained by interferometry \citep{Dong2019,Cassan2021}, which may open very interesting perspectives for very bright sources. Finally, the astrometric detection of the centroid motion provides an alternative channel for space missions \citep{Kluter2020,Sahu2022,Lam2022}.

\subsection{The microlensing contribution to the understanding of Brown Dwarfs}

There are more than 3000 BDs discovered till now: many of them are in the solar neighborhood \citep{Meisner2020}, some are discovered in young clusters \citep{Miret-Roig2021} and some in binary systems. For FGK stars the absence of BDs in a close orbit <5AU has led to the postulation of a BD desert \citep{Grether2006}. Our lens OGLE-2019-BLG-0033 consists of a low-mass M-dwarf as a primary and a BD as a companion with projected separation of $0.587$ au. There are many theories explaining the formation of BD binaries: \citet{Stella2010} describes the formation of low-mass binaries via turbulent fragmentation with separations up to $10^4$ au.  \citet{Fontanive2019} state that there should be a wide-orbit companion for a low mass star having a close-in orbiting BD; such wide orbit companion would have a central role in the formation of the BD and also for the sparse population of BDs in close-in orbits \citep{Irwin2010}. Since low-mass binaries are difficult to observe directly, microlensing will play an increasingly important role in identifying such systems, measuring the masses of BDs in binaries and quantifying their occurrence throughout the Galaxy. Kinematic studies combining relative lens-source proper motions from microlensing and source proper motion from Gaia give very interesting perspectives for assigning low-mass systems to the correct dynamical component of the Galaxy and understanding how the production of BDs may have evolved during the history of the Milky Way.

One of the goals of the upcoming {\it Roman} Galactic Exoplanet Survey (RGES) is the determination of the planetary mass function at $10\%$ in decades \citep{Penny2019,Johnson2020}. By combining high-resolution imaging, space parallax and precise characterization of resolved sources, this space mission is likely to provide a substantial census of BDs, both isolated ones and those in binary systems. Some additional detections are also expected by the xallarap effect \citep{Miyazaki2021}. In order to exploit all this potential, it is necessary to pay adequate attention to equal-mass binary-lens events, even if it is clear that they do not lead to the discovery of planets. 

Compared to the BD science from the {\it James Webb Space Telescope} (JWST) \citep{Ryan2016}, {\it Roman} has a 100 times wider field of view, enabling the possibility of direct detection of a great number of BDs as a byproduct of its survey operations as a whole, in addition to those that will be detected through microlensing. JWST will be instead well-suited for the detection of BDs and rogue planet search in smaller fields, such as clusters, and for the detailed investigation of nearby BDs.

Finally, in a longer perspective, the Extremely Large Telescope, equipped with advanced Adaptive Optics \citep{Trippe2010}, will be able to provide exquisite astrometry in crowded fields. This would be an extraordinary opportunity to revisit all past microlensing events. In fact, current microlensing surveys discover about 100 microlensing binary events every year, with most of them likely composed of low-mass objects. A systematic astrometric investigation of all events would thus build a very broad, detailed and reliable statistics of binary systems in our Galaxy. 

\section{Conclusions}

We have presented the full analysis of OGLE-2019-BLG-0033, a microlensing event discovered by the OGLE survey and observed by many ground telescopes and from the {\it Spitzer} spacecraft. The event is long-enough to allow accurate parallax and orbital motion measurements, along with a detailed characterization of the bright background source. With these favorable circumstances, we manage to achieve an exceptionally precise mass measurement for the lens system, which turns out to be composed of a $0.149 M_\sun$ red dwarf and a Brown Dwarf of $0.0463 M_\sun$ at a projected separation of $0.585$ au. The precision of this mass measurement is $6.8\%$, which is one of the best ever achieved in microlensing observations. The kinematic analysis shows that this binary system is part of the old metal-poor thick disk component of our Galaxy. We argue that the upcoming {\it Roman} Galactic Exoplanet Survey will represent a major advance in our understanding of any classes of sub-stellar objects.

\begin{acknowledgements}
The MOA project is supported by JSPS KAK-ENHI Grant Number JSPS24253004, JSPS26247023, JSPS23340064, JSPS15H00781, JP16H06287,17H02871 and 19KK0082 
and a Royal Society of New Zealand Marsden Grant MFP-MAU1901.

U.G.J. acknowledges funding from the European Union H2020-MSCA-ITN-2019 under Grant no. 860470 (CHAMELEON) and from the Novo Nordisk Foundation Interdisciplinary Synergy Program grant no. NNF19OC0057374.

W.Zang, W.Zhu, H.Y. and S.M. acknowledge support by the National Science Foundation of China (Grant No. 12133005). This research uses data obtained through the Telescope Access Program (TAP), which has been funded by the TAP member institutes.

Work by C.R. was supported by a Research Fellowship of the Alexander von Humboldt Foundation.

Work by C.H. was supported by the grants of National Research
Foundation of Korea (2020R1A4A2002885 and 2019R1A2C2085965).

Work by JCY was supported by JPL grant 1571564.

This work has made use of data from the European Space Agency (ESA) mission
{\it Gaia} (\url{https://www.cosmos.esa.int/gaia}), processed by the {\it Gaia}
Data Processing and Analysis Consortium (DPAC,
\url{https://www.cosmos.esa.int/web/gaia/dpac/consortium}). Funding for the DPAC
has been provided by national institutions, in particular the institutions
participating in the {\it Gaia} Multilateral Agreement.

This work uses observations made at the Observatório do Pico dos
Dias/LNA (Brazil).

\end{acknowledgements}

%
\bibliographystyle{aa} 
\bibliography{bibliografia} 

\begin{thebibliography}{108}
\expandafter\ifx\csname natexlab\endcsname\relax\def\natexlab#1{#1}\fi

\bibitem[{{Alard}(2000)}]{Alard2000}
{Alard}, C. 2000, \aaps, 144, 363

\bibitem[{{Alard} \& {Lupton}(1998)}]{Alard1998}
{Alard}, C. \& {Lupton}, R.~H. 1998, \apj, 503, 325

\bibitem[{{Albrow} {et~al.}(2018){Albrow}, {Yee}, {Udalski}, {Calchi Novati},
  {Carey}, {Henderson}, {Beichman}, {Bryden}, {Gaudi}, {Shvartzvald}, {Spitzer
  Team}, {Szyma{\'n}ski}, {Mr{\'o}z}, {Skowron}, {Poleski}, {Soszy{\'n}ski},
  {Koz{\l}owski}, {Pietrukowicz}, {Ulaczyk}, {Pawlak}, {OGLE Collaboration},
  {Chung}, {Gould}, {Han}, {Hwang}, {Jung}, {Ryu}, {Shin}, {Zhu}, {Cha}, {Kim},
  {Kim}, {Kim}, {Lee}, {Lee}, {Lee}, {Park}, {Pogge}, \& {KMTNet
  Collaboration}}]{Albrow2018}
{Albrow}, M.~D., {Yee}, J.~C., {Udalski}, A., {et~al.} 2018, \apj, 858, 107

\bibitem[{{An} {et~al.}(2002){An}, {Albrow}, {Beaulieu}, {Caldwell}, {DePoy},
  {Dominik}, {Gaudi}, {Gould}, {Greenhill}, {Hill}, {Kane}, {Martin},
  {Menzies}, {Pogge}, {Pollard}, {Sackett}, {Sahu}, {Vermaak}, {Watson}, \&
  {Williams}}]{An2002}
{An}, J.~H., {Albrow}, M.~D., {Beaulieu}, J.~P., {et~al.} 2002, \apj, 572, 521

\bibitem[{{Aparicio} \& {Gallart}(2004)}]{Aparicio2004}
{Aparicio}, A. \& {Gallart}, C. 2004, \aj, 128, 1465

\bibitem[{{Auddy} {et~al.}(2016){Auddy}, {Basu}, \& {Valluri}}]{Auddy2016}
{Auddy}, S., {Basu}, S., \& {Valluri}, S.~R. 2016, Advances in Astronomy, 2016,
  574327

\bibitem[{{Batista} {et~al.}(2011){Batista}, {Gould}, {Dieters}, {Dong},
  {Bond}, {Beaulieu}, {Maoz}, {Monard}, {Christie}, {McCormick}, {Albrow},
  {Horne}, {Tsapras}, {Burgdorf}, {Calchi Novati}, {Skottfelt}, {Caldwell},
  {Koz{\l}owski}, {Kubas}, {Gaudi}, {Han}, {Bennett}, {An}, {MOA
  Collaboration}, {Abe}, {Botzler}, {Douchin}, {Freeman}, {Fukui}, {Furusawa},
  {Hearnshaw}, {Hosaka}, {Itow}, {Kamiya}, {Kilmartin}, {Korpela}, {Lin},
  {Ling}, {Makita}, {Masuda}, {Matsubara}, {Miyake}, {Muraki}, {Nagaya},
  {Nishimoto}, {Ohnishi}, {Okumura}, {Perrott}, {Rattenbury}, {Saito},
  {Sullivan}, {Sumi}, {Sweatman}, {Tristram}, {von Seggern}, {Yock}, {PLANET
  Collaboration}, {Brillant}, {Calitz}, {Cassan}, {Cole}, {Cook}, {Coutures},
  {Dominis Prester}, {Donatowicz}, {Greenhill}, {Hoffman}, {Jablonski}, {Kane},
  {Kains}, {Marquette}, {Martin}, {Martioli}, {Meintjes}, {Menzies},
  {Pedretti}, {Pollard}, {Sahu}, {Vinter}, {Wambsganss}, {Watson}, {Williams},
  {Zub}, {FUN Collaboration}, {Allen}, {Bolt}, {Bos}, {DePoy}, {Drummond},
  {Eastman}, {Gal-Yam}, {Gorbikov}, {Higgins}, {Janczak}, {Kaspi}, {Lee},
  {Mallia}, {Maury}, {Monard}, {Moorhouse}, {Morgan}, {Natusch}, {Ofek},
  {Park}, {Pogge}, {Polishook}, {Santallo}, {Shporer}, {Spector}, {Thornley},
  {Yee}, {MiNDSTEp Consortium}, {Bozza}, {Browne}, {Dominik}, {Dreizler},
  {Finet}, {Glitrup}, {Grundahl}, {Harps{\o}e}, {Hessman}, {Hinse},
  {Hundertmark}, {J{\o}rgensen}, {Liebig}, {Maier}, {Mancini}, {Mathiasen},
  {Rahvar}, {Ricci}, {Scarpetta}, {Southworth}, {Surdej}, {Zimmer}, {RoboNet
  Collaboration}, {Allan}, {Bramich}, {Snodgrass}, {Steele}, \&
  {Street}}]{Batista2011}
{Batista}, V., {Gould}, A., {Dieters}, S., {et~al.} 2011, \aap, 529, A102

\bibitem[{{B{\'e}jar} {et~al.}(2001){B{\'e}jar}, {Mart{\'\i}n}, {Zapatero
  Osorio}, {Rebolo}, {Barrado y Navascu{\'e}s}, {Bailer-Jones}, {Mundt},
  {Baraffe}, {Chabrier}, \& {Allard}}]{Bejar2001}
{B{\'e}jar}, V.~J.~S., {Mart{\'\i}n}, E.~L., {Zapatero Osorio}, M.~R., {et~al.}
  2001, \apj, 556, 830

\bibitem[{{Benedict} {et~al.}(2016){Benedict}, {Henry}, {Franz}, {McArthur},
  {Wasserman}, {Jao}, {Cargile}, {Dieterich}, {Bradley}, {Nelan}, \&
  {Whipple}}]{Benedict2016}
{Benedict}, G.~F., {Henry}, T.~J., {Franz}, O.~G., {et~al.} 2016, \aj, 152, 141

\bibitem[{{Bennett} {et~al.}(2008){Bennett}, {Bond}, {Udalski}, {Sumi}, {Abe},
  {Fukui}, {Furusawa}, {Hearnshaw}, {Holderness}, {Itow}, {Kamiya}, {Korpela},
  {Kilmartin}, {Lin}, {Ling}, {Masuda}, {Matsubara}, {Miyake}, {Muraki},
  {Nagaya}, {Okumura}, {Ohnishi}, {Perrott}, {Rattenbury}, {Sako}, {Saito},
  {Sato}, {Skuljan}, {Sullivan}, {Sweatman}, {Tristram}, {Yock}, {Kubiak},
  {Szyma{\'n}ski}, {Pietrzy{\'n}ski}, {Soszy{\'n}ski}, {Szewczyk},
  {Wyrzykowski}, {Ulaczyk}, {Batista}, {Beaulieu}, {Brillant}, {Cassan},
  {Fouqu{\'e}}, {Kervella}, {Kubas}, \& {Marquette}}]{Bennett2008}
{Bennett}, D.~P., {Bond}, I.~A., {Udalski}, A., {et~al.} 2008, \apj, 684, 663

\bibitem[{{Bensby} {et~al.}(2011){Bensby}, {Ad{\'e}n}, {Mel{\'e}ndez}, {Gould},
  {Feltzing}, {Asplund}, {Johnson}, {Lucatello}, {Yee}, {Ram{\'\i}rez},
  {Cohen}, {Thompson}, {Bond}, {Gal-Yam}, {Han}, {Sumi}, {Suzuki}, {Wada},
  {Miyake}, {Furusawa}, {Ohmori}, {Saito}, {Tristram}, \&
  {Bennett}}]{Bensby2011}
{Bensby}, T., {Ad{\'e}n}, D., {Mel{\'e}ndez}, J., {et~al.} 2011, \aap, 533,
  A134

\bibitem[{{Bensby} {et~al.}(2010){Bensby}, {Feltzing}, {Johnson}, {Gould},
  {Ad{\'e}n}, {Asplund}, {Mel{\'e}ndez}, {Gal-Yam}, {Lucatello}, {Sana},
  {Sumi}, {Miyake}, {Suzuki}, {Han}, {Bond}, \& {Udalski}}]{Bensby2010}
{Bensby}, T., {Feltzing}, S., {Johnson}, J.~A., {et~al.} 2010, \aap, 512, A41

\bibitem[{{Bensby} {et~al.}(2021){Bensby}, {Gould}, {Asplund}, {Feltzing},
  {Mel{\'e}ndez}, {Johnson}, {Lucatello}, {Udalski}, \& {Yee}}]{Bensby2021}
{Bensby}, T., {Gould}, A., {Asplund}, M., {et~al.} 2021, \aap, 655, A117

\bibitem[{{Bertelli} {et~al.}(1994){Bertelli}, {Bressan}, {Chiosi}, {Fagotto},
  \& {Nasi}}]{Bertelli1994}
{Bertelli}, G., {Bressan}, A., {Chiosi}, C., {Fagotto}, F., \& {Nasi}, E. 1994,
  \aaps, 106, 275

\bibitem[{{Bessell} \& {Brett}(1988)}]{Bessel1988}
{Bessell}, M.~S. \& {Brett}, J.~M. 1988, \pasp, 100, 1134

\bibitem[{{Bhattacharya} {et~al.}(2018){Bhattacharya}, {Beaulieu}, {Bennett},
  {Anderson}, {Koshimoto}, {Lu}, {Batista}, {Blackman}, {Bond}, {Fukui},
  {Henderson}, {Hirao}, {Marquette}, {Mroz}, {Ranc}, \&
  {Udalski}}]{Bhattacharya2018}
{Bhattacharya}, A., {Beaulieu}, J.~P., {Bennett}, D.~P., {et~al.} 2018, \aj,
  156, 289

\bibitem[{{Blackman} {et~al.}(2021){Blackman}, {Beaulieu}, {Bennett},
  {Danielski}, {Alard}, {Cole}, {Vandorou}, {Ranc}, {Terry}, {Bhattacharya},
  {Bond}, {Bachelet}, {Veras}, {Koshimoto}, {Batista}, \&
  {Marquette}}]{Blackman2021}
{Blackman}, J.~W., {Beaulieu}, J.~P., {Bennett}, D.~P., {et~al.} 2021, \nat,
  598, 272

\bibitem[{{Bond} {et~al.}(2001){Bond}, {Abe}, {Dodd}, {Hearnshaw}, {Honda},
  {Jugaku}, {Kilmartin}, {Marles}, {Masuda}, {Matsubara}, {Muraki}, {Nakamura},
  {Nankivell}, {Noda}, {Noguchi}, {Ohnishi}, {Rattenbury}, {Reid}, {Saito},
  {Sato}, {Sekiguchi}, {Skuljan}, {Sullivan}, {Sumi}, {Takeuti}, {Watase},
  {Wilkinson}, {Yamada}, {Yanagisawa}, \& {Yock}}]{Bond2001}
{Bond}, I.~A., {Abe}, F., {Dodd}, R.~J., {et~al.} 2001, \mnras, 327, 868

\bibitem[{{Bond} {et~al.}(2017){Bond}, {Bennett}, {Sumi}, {Udalski}, {Suzuki},
  {Rattenbury}, {Bozza}, {Koshimoto}, {Abe}, {Asakura}, {Barry},
  {Bhattacharya}, {Donachie}, {Evans}, {Fukui}, {Hirao}, {Itow}, {Li}, {Ling},
  {Masuda}, {Matsubara}, {Muraki}, {Nagakane}, {Ohnishi}, {Ranc}, {Saito},
  {Sharan}, {Sullivan}, {Tristram}, {Yamada}, {Yamada}, {Yonehara}, {Skowron},
  {Szyma{\'n}ski}, {Poleski}, {Mr{\'o}z}, {Soszy{\'n}ski}, {Pietrukowicz},
  {Koz{\l}owski}, {Ulaczyk}, \& {Pawlak}}]{Bond2017}
{Bond}, I.~A., {Bennett}, D.~P., {Sumi}, T., {et~al.} 2017, \mnras, 469, 2434

\bibitem[{{Bozza}(2000)}]{Bozza2000}
{Bozza}, V. 2000, \aap, 355, 423

\bibitem[{{Bozza}(2010)}]{Bozza2010}
{Bozza}, V. 2010, \mnras, 408, 2188

\bibitem[{{Bozza} {et~al.}(2018){Bozza}, {Bachelet}, {Bartoli{\'c}}, {Heintz},
  {Hoag}, \& {Hundertmark}}]{Bozza2018}
{Bozza}, V., {Bachelet}, E., {Bartoli{\'c}}, F., {et~al.} 2018, \mnras, 479,
  5157

\bibitem[{{Bozza} {et~al.}(2012){Bozza}, {Dominik}, {Rattenbury},
  {J{\o}rgensen}, {Tsapras}, {Bramich}, {Udalski}, {Bond}, {Liebig}, {Cassan},
  {Fouqu{\'e}}, {Fukui}, {Hundertmark}, {Shin}, {Lee}, {Choi}, {Park}, {Gould},
  {Allan}, {Mao}, {Wyrzykowski}, {Street}, {Buckley}, {Nagayama}, {Mathiasen},
  {Hinse}, {Novati}, {Harps{\o}e}, {Mancini}, {Scarpetta}, {Anguita},
  {Burgdorf}, {Horne}, {Hornstrup}, {Kains}, {Kerins}, {Kj{\ae}rgaard}, {Masi},
  {Rahvar}, {Ricci}, {Snodgrass}, {Southworth}, {Steele}, {Surdej},
  {Th{\"o}ne}, {Wambsganss}, {Zub}, {Albrow}, {Batista}, {Beaulieu}, {Bennett},
  {Caldwell}, {Cole}, {Cook}, {Coutures}, {Dieters}, {Dominis Prester},
  {Donatowicz}, {Greenhill}, {Kane}, {Kubas}, {Marquette}, {Martin}, {Menzies},
  {Pollard}, {Sahu}, {Williams}, {Szyma{\'n}ski}, {Kubiak}, {Pietrzy{\'n}ski},
  {Soszy{\'n}ski}, {Poleski}, {Ulaczyk}, {DePoy}, {Dong}, {Han}, {Janczak},
  {Lee}, {Pogge}, {Abe}, {Furusawa}, {Hearnshaw}, {Itow}, {Kilmartin},
  {Korpela}, {Lin}, {Ling}, {Masuda}, {Matsubara}, {Miyake}, {Muraki},
  {Ohnishi}, {Perrott}, {Saito}, {Skuljan}, {Sullivan}, {Sumi}, {Suzuki},
  {Sweatman}, {Tristram}, {Wada}, {Yock}, {Gulbis}, {Hashimoto}, {Kniazev}, \&
  {Vaisanen}}]{Bozza2012}
{Bozza}, V., {Dominik}, M., {Rattenbury}, N.~J., {et~al.} 2012, \mnras, 424,
  902

\bibitem[{{Bozza} {et~al.}(2021){Bozza}, {Khalouei}, \& {Bachelet}}]{Bozza2021}
{Bozza}, V., {Khalouei}, E., \& {Bachelet}, E. 2021, \mnras, 505, 126

\bibitem[{{Bramich}(2008)}]{Bramich2008}
{Bramich}, D.~M. 2008, \mnras, 386, L77

\bibitem[{{Bramich} {et~al.}(2013){Bramich}, {Horne}, {Albrow}, {Tsapras},
  {Snodgrass}, {Street}, {Hundertmark}, {Kains}, {Arellano Ferro}, {Figuera},
  \& {Giridhar}}]{Bramich2013}
{Bramich}, D.~M., {Horne}, K., {Albrow}, M.~D., {et~al.} 2013, \mnras, 428,
  2275

\bibitem[{{Burrows} {et~al.}(1997){Burrows}, {Marley}, {Hubbard}, {Lunine},
  {Guillot}, {Saumon}, {Freedman}, {Sudarsky}, \& {Sharp}}]{Burrows1997}
{Burrows}, A., {Marley}, M., {Hubbard}, W.~B., {et~al.} 1997, \apj, 491, 856

\bibitem[{{Calchi Novati} {et~al.}(2015){Calchi Novati}, {Gould}, {Yee},
  {Beichman}, {Bryden}, {Carey}, {Fausnaugh}, {Gaudi}, {Henderson}, {Pogge},
  {Shvartzvald}, {Wibking}, {Zhu}, {Spitzer Team}, {Udalski}, {Poleski},
  {Pawlak}, {Szyma{\'n}ski}, {Skowron}, {Mr{\'o}z}, {Koz{\l}owski},
  {Wyrzykowski}, {Pietrukowicz}, {Pietrzy{\'n}ski}, {Soszy{\'n}ski}, {Ulaczyk},
  \& {OGLE Group}}]{Calchi2015}
{Calchi Novati}, S., {Gould}, A., {Yee}, J.~C., {et~al.} 2015, \apj, 814, 92

\bibitem[{{Cassan} {et~al.}(2021){Cassan}, {Ranc}, {Absil}, {Wyrzykowski},
  {Rybicki}, {Bachelet}, {Le Bouquin}, {Hundertmark}, {Street}, {Surdej},
  {Tsapras}, {Wambsganss}, \& {Wertz}}]{Cassan2021}
{Cassan}, A., {Ranc}, C., {Absil}, O., {et~al.} 2021, Nature Astronomy

\bibitem[{{Castelli} \& {Kurucz}(2003)}]{Castelli2003}
{Castelli}, F. \& {Kurucz}, R.~L. 2003, in Modelling of Stellar Atmospheres,
  ed. N.~{Piskunov}, W.~W. {Weiss}, \& D.~F. {Gray}, Vol. 210, A20

\bibitem[{{Choi} {et~al.}(2013){Choi}, {Han}, {Udalski}, {Sumi}, {Gaudi},
  {Gould}, {Bennett}, {Dominik}, {Beaulieu}, {Tsapras}, {Bozza}, {Abe}, {Bond},
  {Botzler}, {Chote}, {Freeman}, {Fukui}, {Furusawa}, {Itow}, {Ling}, {Masuda},
  {Matsubara}, {Miyake}, {Muraki}, {Ohnishi}, {Rattenbury}, {Saito},
  {Sullivan}, {Suzuki}, {Sweatman}, {Suzuki}, {Takino}, {Tristram}, {Wada},
  {Yock}, {MOA Collaboration}, {Szyma{\'n}ski}, {Kubiak}, {Pietrzy{\'n}ski},
  {Soszy{\'n}ski}, {Skowron}, {Koz{\l}owski}, {Poleski}, {Ulaczyk},
  {Wyrzykowski}, {Pietrukowicz}, {OGLE Collaboration}, {Almeida}, {DePoy},
  {Dong}, {Gorbikov}, {Jablonski}, {Henderson}, {Hwang}, {Janczak}, {Jung},
  {Kaspi}, {Lee}, {Malamud}, {Maoz}, {McGregor}, {Mu{\~n}oz}, {Park}, {Park},
  {Pogge}, {Shvartzvald}, {Shin}, {Yee}, {{\ensuremath{\mu}}FUN Collaboration},
  {Alsubai}, {Browne}, {Burgdorf}, {Calchi Novati}, {Dodds}, {Fang}, {Finet},
  {Glitrup}, {Grundahl}, {Gu}, {Hardis}, {Harps{\o}e}, {Hinse}, {Hornstrup},
  {Hundertmark}, {Jessen-Hansen}, {Jrgensen}, {Kains}, {Kerins}, {Liebig},
  {Lund}, {Lundkvist}, {Maier}, {Mancini}, {Mathiasen}, {Penny}, {Rahvar},
  {Ricci}, {Scarpetta}, {Skottfelt}, {Snodgrass}, {Southworth}, {Surdej},
  {Tregloan-Reed}, {Wambsganss}, {Wertz}, {Zimmer}, {MiNDSTEp Consortium},
  {Albrow}, {Bachelet}, {Batista}, {Brillant}, {Cassan}, {Cole}, {Coutures},
  {Dieters}, {Dominis Prester}, {Donatowicz}, {Fouqu{\'e}}, {Greenhill},
  {Kubas}, {Marquette}, {Menzies}, {Sahu}, {Zub}, {PLANET Collaboration},
  {Bramich}, {Horne}, {Steele}, {Street}, \& {RoboNet
  Collaboration}}]{Choi2013}
{Choi}, J.~Y., {Han}, C., {Udalski}, A., {et~al.} 2013, \apj, 768, 129

\bibitem[{{Claret} \& {Bloemen}(2011)}]{Claret2011}
{Claret}, A. \& {Bloemen}, S. 2011, \aap, 529, A75

\bibitem[{{Close} {et~al.}(2007){Close}, {Zuckerman}, {Song}, {Barman},
  {Marois}, {Rice}, {Siegler}, {Macintosh}, {Becklin}, {Campbell}, {Lyke},
  {Conrad}, \& {Le Mignant}}]{close2007}
{Close}, L.~M., {Zuckerman}, B., {Song}, I., {et~al.} 2007, \apj, 660, 1492

\bibitem[{{Dominik}(1999)}]{Dominik1999}
{Dominik}, M. 1999, \aap, 349, 108

\bibitem[{{Dominik}(2006)}]{Dominik2006}
{Dominik}, M. 2006, \mnras, 367, 669

\bibitem[{{Dominik} {et~al.}(2010){Dominik}, {J{\o}rgensen}, {Rattenbury},
  {Mathiasen}, {Hinse}, {Calchi Novati}, {Harps{\o}e}, {Bozza}, {Anguita},
  {Burgdorf}, {Horne}, {Hundertmark}, {Kerins}, {Kj{\ae}rgaard}, {Liebig},
  {Mancini}, {Masi}, {Rahvar}, {Ricci}, {Scarpetta}, {Snodgrass}, {Southworth},
  {Street}, {Surdej}, {Th{\"o}ne}, {Tsapras}, {Wambsganss}, \&
  {Zub}}]{Dominik2010}
{Dominik}, M., {J{\o}rgensen}, U.~G., {Rattenbury}, N.~J., {et~al.} 2010,
  Astronomische Nachrichten, 331, 671

\bibitem[{{Dong} {et~al.}(2019){Dong}, {M{\'e}rand}, {Delplancke-Str{\"o}bele},
  {Gould}, {Chen}, {Post}, {Kochanek}, {Stanek}, {Christie}, {Mutel},
  {Natusch}, {Holoien}, {Prieto}, {Shappee}, \& {Thompson}}]{Dong2019}
{Dong}, S., {M{\'e}rand}, A., {Delplancke-Str{\"o}bele}, F., {et~al.} 2019,
  \apj, 871, 70

\bibitem[{{Elmegreen}(1997)}]{Elmegreen1997}
{Elmegreen}, B.~G. 1997, \apj, 486, 944

\bibitem[{{Fontanive} {et~al.}(2019){Fontanive}, {Rice}, {Bonavita}, {Lopez},
  {Mu{\v{z}}i{\'c}}, {}, \& {Biller}}]{Fontanive2019}
{Fontanive}, C., {Rice}, K., {Bonavita}, M., {et~al.} 2019, \mnras, 485, 4967

\bibitem[{{Gaia Collaboration} {et~al.}(2021){Gaia Collaboration}, {Brown},
  {Vallenari}, {Prusti}, {de Bruijne}, {Babusiaux}, {Biermann}, {Creevey},
  {Evans}, {Eyer}, {Hutton}, {Jansen}, {Jordi}, {Klioner}, {Lammers},
  {Lindegren}, {Luri}, {Mignard}, {Panem}, {Pourbaix}, {Randich}, {Sartoretti},
  {Soubiran}, {Walton}, {Arenou}, {Bailer-Jones}, {Bastian}, {Cropper},
  {Drimmel}, {Katz}, {Lattanzi}, {van Leeuwen}, {Bakker}, {Cacciari},
  {Casta{\~n}eda}, {De Angeli}, {Ducourant}, {Fabricius}, {Fouesneau},
  {Fr{\'e}mat}, {Guerra}, {Guerrier}, {Guiraud}, {Jean-Antoine Piccolo},
  {Masana}, {Messineo}, {Mowlavi}, {Nicolas}, {Nienartowicz}, {Pailler},
  {Panuzzo}, {Riclet}, {Roux}, {Seabroke}, {Sordo}, {Tanga}, {Th{\'e}venin},
  {Gracia-Abril}, {Portell}, {Teyssier}, {Altmann}, {Andrae}, {Bellas-Velidis},
  {Benson}, {Berthier}, {Blomme}, {Brugaletta}, {Burgess}, {Busso}, {Carry},
  {Cellino}, {Cheek}, {Clementini}, {Damerdji}, {Davidson}, {Delchambre},
  {Dell'Oro}, {Fern{\'a}ndez-Hern{\'a}ndez}, {Galluccio}, {Garc{\'\i}a-Lario},
  {Garcia-Reinaldos}, {Gonz{\'a}lez-N{\'u}{\~n}ez}, {Gosset}, {Haigron},
  {Halbwachs}, {Hambly}, {Harrison}, {Hatzidimitriou}, {Heiter},
  {Hern{\'a}ndez}, {Hestroffer}, {Hodgkin}, {Holl}, {Jan{\ss}en}, {Jevardat de
  Fombelle}, {Jordan}, {Krone-Martins}, {Lanzafame}, {L{\"o}ffler}, {Lorca},
  {Manteiga}, {Marchal}, {Marrese}, {Moitinho}, {Mora}, {Muinonen}, {Osborne},
  {Pancino}, {Pauwels}, {Petit}, {Recio-Blanco}, {Richards}, {Riello},
  {Rimoldini}, {Robin}, {Roegiers}, {Rybizki}, {Sarro}, {Siopis}, {Smith},
  {Sozzetti}, {Ulla}, {Utrilla}, {van Leeuwen}, {van Reeven}, {Abbas}, {Abreu
  Aramburu}, {Accart}, {Aerts}, {Aguado}, {Ajaj}, {Altavilla}, {{\'A}lvarez},
  {{\'A}lvarez Cid-Fuentes}, {Alves}, {Anderson}, {Anglada Varela}, {Antoja},
  {Audard}, {Baines}, {Baker}, {Balaguer-N{\'u}{\~n}ez}, {Balbinot}, {Balog},
  {Barache}, {Barbato}, {Barros}, {Barstow}, {Bartolom{\'e}}, {Bassilana},
  {Bauchet}, {Baudesson-Stella}, {Becciani}, {Bellazzini}, {Bernet}, {Bertone},
  {Bianchi}, {Blanco-Cuaresma}, {Boch}, {Bombrun}, {Bossini}, {Bouquillon},
  {Bragaglia}, {Bramante}, {Breedt}, {Bressan}, {Brouillet}, {Bucciarelli},
  {Burlacu}, {Busonero}, {Butkevich}, {Buzzi}, {Caffau}, {Cancelliere},
  {C{\'a}novas}, {Cantat-Gaudin}, {Carballo}, {Carlucci}, {Carnerero},
  {Carrasco}, {Casamiquela}, {Castellani}, {Castro-Ginard}, {Castro Sampol},
  {Chaoul}, {Charlot}, {Chemin}, {Chiavassa}, {Cioni}, {Comoretto}, {Cooper},
  {Cornez}, {Cowell}, {Crifo}, {Crosta}, {Crowley}, {Dafonte}, {Dapergolas},
  {David}, {David}, {de Laverny}, {De Luise}, {De March}, {De Ridder}, {de
  Souza}, {de Teodoro}, {de Torres}, {del Peloso}, {del Pozo}, {Delbo},
  {Delgado}, {Delgado}, {Delisle}, {Di Matteo}, {Diakite}, {Diener},
  {Distefano}, {Dolding}, {Eappachen}, {Edvardsson}, {Enke}, {Esquej}, {Fabre},
  {Fabrizio}, {Faigler}, {Fedorets}, {Fernique}, {Fienga}, {Figueras},
  {Fouron}, {Fragkoudi}, {Fraile}, {Franke}, {Gai}, {Garabato},
  {Garcia-Gutierrez}, {Garc{\'\i}a-Torres}, {Garofalo}, {Gavras}, {Gerlach},
  {Geyer}, {Giacobbe}, {Gilmore}, {Girona}, {Giuffrida}, {Gomel}, {Gomez},
  {Gonzalez-Santamaria}, {Gonz{\'a}lez-Vidal}, {Granvik},
  {Guti{\'e}rrez-S{\'a}nchez}, {Guy}, {Hauser}, {Haywood}, {Helmi}, {Hidalgo},
  {Hilger}, {H{\l}adczuk}, {Hobbs}, {Holland}, {Huckle}, {Jasniewicz},
  {Jonker}, {Juaristi Campillo}, {Julbe}, {Karbevska}, {Kervella}, {Khanna},
  {Kochoska}, {Kontizas}, {Kordopatis}, {Korn}, {Kostrzewa-Rutkowska},
  {Kruszy{\'n}ska}, {Lambert}, {Lanza}, {Lasne}, {Le Campion}, {Le Fustec},
  {Lebreton}, {Lebzelter}, {Leccia}, {Leclerc}, {Lecoeur-Taibi}, {Liao},
  {Licata}, {Lindstr{\o}m}, {Lister}, {Livanou}, {Lobel}, {Madrero Pardo},
  {Managau}, {Mann}, {Marchant}, {Marconi}, {Marcos Santos}, {Marinoni},
  {Marocco}, {Marshall}, {Martin Polo}, {Mart{\'\i}n-Fleitas}, {Masip},
  {Massari}, {Mastrobuono-Battisti}, {Mazeh}, {McMillan}, {Messina},
  {Michalik}, {Millar}, {Mints}, {Molina}, {Molinaro}, {Moln{\'a}r},
  {Montegriffo}, {Mor}, {Morbidelli}, {Morel}, {Morris}, {Mulone}, {Munoz},
  {Muraveva}, {Murphy}, {Musella}, {Noval}, {Ord{\'e}novic}, {Orr{\`u}},
  {Osinde}, {Pagani}, {Pagano}, {Palaversa}, {Palicio}, {Panahi}, {Pawlak},
  {Pe{\~n}alosa Esteller}, {Penttil{\"a}}, {Piersimoni}, {Pineau}, {Plachy},
  {Plum}, {Poggio}, {Poretti}, {Poujoulet}, {Pr{\v{s}}a}, {Pulone}, {Racero},
  {Ragaini}, {Rainer}, {Raiteri}, {Rambaux}, {Ramos}, {Ramos-Lerate}, {Re
  Fiorentin}, {Regibo}, {Reyl{\'e}}, {Ripepi}, {Riva}, {Rixon}, {Robichon},
  {Robin}, {Roelens}, {Rohrbasser}, {Romero-G{\'o}mez}, {Rowell}, {Royer},
  {Rybicki}, {Sadowski}, {Sagrist{\`a} Sell{\'e}s}, {Sahlmann}, {Salgado},
  {Salguero}, {Samaras}, {Sanchez Gimenez}, {Sanna}, {Santove{\~n}a},
  {Sarasso}, {Schultheis}, {Sciacca}, {Segol}, {Segovia}, {S{\'e}gransan},
  {Semeux}, {Shahaf}, {Siddiqui}, {Siebert}, {Siltala}, {Slezak}, {Smart},
  {Solano}, {Solitro}, {Souami}, {Souchay}, {Spagna}, {Spoto}, {Steele},
  {Steidelm{\"u}ller}, {Stephenson}, {S{\"u}veges}, {Szabados}, {Szegedi-Elek},
  {Taris}, {Tauran}, {Taylor}, {Teixeira}, {Thuillot}, {Tonello}, {Torra},
  {Torra}, {Turon}, {Unger}, {Vaillant}, {van Dillen}, {Vanel}, {Vecchiato},
  {Viala}, {Vicente}, {Voutsinas}, {Weiler}, {Wevers}, {Wyrzykowski}, {Yoldas},
  {Yvard}, {Zhao}, {Zorec}, {Zucker}, {Zurbach}, \& {Zwitter}}]{Gaia2021}
{Gaia Collaboration}, {Brown}, A.~G.~A., {Vallenari}, A., {et~al.} 2021, \aap,
  649, A1

\bibitem[{{Gaia Collaboration} {et~al.}(2016){Gaia Collaboration}, {Prusti},
  {de Bruijne}, {Brown}, {Vallenari}, {Babusiaux}, {Bailer-Jones}, {Bastian},
  {Biermann}, {Evans}, {Eyer}, {Jansen}, {Jordi}, {Klioner}, {Lammers},
  {Lindegren}, {Luri}, {Mignard}, {Milligan}, {Panem}, {Poinsignon},
  {Pourbaix}, {Randich}, {Sarri}, {Sartoretti}, {Siddiqui}, {Soubiran},
  {Valette}, {van Leeuwen}, {Walton}, {Aerts}, {Arenou}, {Cropper}, {Drimmel},
  {H{\o}g}, {Katz}, {Lattanzi}, {O'Mullane}, {Grebel}, {Holland}, {Huc},
  {Passot}, {Bramante}, {Cacciari}, {Casta{\~n}eda}, {Chaoul}, {Cheek}, {De
  Angeli}, {Fabricius}, {Guerra}, {Hern{\'a}ndez}, {Jean-Antoine-Piccolo},
  {Masana}, {Messineo}, {Mowlavi}, {Nienartowicz}, {Ord{\'o}{\~n}ez-Blanco},
  {Panuzzo}, {Portell}, {Richards}, {Riello}, {Seabroke}, {Tanga},
  {Th{\'e}venin}, {Torra}, {Els}, {Gracia-Abril}, {Comoretto},
  {Garcia-Reinaldos}, {Lock}, {Mercier}, {Altmann}, {Andrae}, {Astraatmadja},
  {Bellas-Velidis}, {Benson}, {Berthier}, {Blomme}, {Busso}, {Carry},
  {Cellino}, {Clementini}, {Cowell}, {Creevey}, {Cuypers}, {Davidson}, {De
  Ridder}, {de Torres}, {Delchambre}, {Dell'Oro}, {Ducourant}, {Fr{\'e}mat},
  {Garc{\'\i}a-Torres}, {Gosset}, {Halbwachs}, {Hambly}, {Harrison}, {Hauser},
  {Hestroffer}, {Hodgkin}, {Huckle}, {Hutton}, {Jasniewicz}, {Jordan},
  {Kontizas}, {Korn}, {Lanzafame}, {Manteiga}, {Moitinho}, {Muinonen},
  {Osinde}, {Pancino}, {Pauwels}, {Petit}, {Recio-Blanco}, {Robin}, {Sarro},
  {Siopis}, {Smith}, {Smith}, {Sozzetti}, {Thuillot}, {van Reeven}, {Viala},
  {Abbas}, {Abreu Aramburu}, {Accart}, {Aguado}, {Allan}, {Allasia},
  {Altavilla}, {{\'A}lvarez}, {Alves}, {Anderson}, {Andrei}, {Anglada Varela},
  {Antiche}, {Antoja}, {Ant{\'o}n}, {Arcay}, {Atzei}, {Ayache}, {Bach},
  {Baker}, {Balaguer-N{\'u}{\~n}ez}, {Barache}, {Barata}, {Barbier}, {Barblan},
  {Baroni}, {Barrado y Navascu{\'e}s}, {Barros}, {Barstow}, {Becciani},
  {Bellazzini}, {Bellei}, {Bello Garc{\'\i}a}, {Belokurov}, {Bendjoya},
  {Berihuete}, {Bianchi}, {Bienaym{\'e}}, {Billebaud}, {Blagorodnova},
  {Blanco-Cuaresma}, {Boch}, {Bombrun}, {Borrachero}, {Bouquillon}, {Bourda},
  {Bouy}, {Bragaglia}, {Breddels}, {Brouillet}, {Br{\"u}semeister},
  {Bucciarelli}, {Budnik}, {Burgess}, {Burgon}, {Burlacu}, {Busonero}, {Buzzi},
  {Caffau}, {Cambras}, {Campbell}, {Cancelliere}, {Cantat-Gaudin}, {Carlucci},
  {Carrasco}, {Castellani}, {Charlot}, {Charnas}, {Charvet}, {Chassat},
  {Chiavassa}, {Clotet}, {Cocozza}, {Collins}, {Collins}, {Costigan}, {Crifo},
  {Cross}, {Crosta}, {Crowley}, {Dafonte}, {Damerdji}, {Dapergolas}, {David},
  {David}, {De Cat}, {de Felice}, {de Laverny}, {De Luise}, {De March}, {de
  Martino}, {de Souza}, {Debosscher}, {del Pozo}, {Delbo}, {Delgado},
  {Delgado}, {di Marco}, {Di Matteo}, {Diakite}, {Distefano}, {Dolding}, {Dos
  Anjos}, {Drazinos}, {Dur{\'a}n}, {Dzigan}, {Ecale}, {Edvardsson}, {Enke},
  {Erdmann}, {Escolar}, {Espina}, {Evans}, {Eynard Bontemps}, {Fabre},
  {Fabrizio}, {Faigler}, {Falc{\~a}o}, {Farr{\`a}s Casas}, {Faye}, {Federici},
  {Fedorets}, {Fern{\'a}ndez-Hern{\'a}ndez}, {Fernique}, {Fienga}, {Figueras},
  {Filippi}, {Findeisen}, {Fonti}, {Fouesneau}, {Fraile}, {Fraser}, {Fuchs},
  {Furnell}, {Gai}, {Galleti}, {Galluccio}, {Garabato}, {Garc{\'\i}a-Sedano},
  {Gar{\'e}}, {Garofalo}, {Garralda}, {Gavras}, {Gerssen}, {Geyer}, {Gilmore},
  {Girona}, {Giuffrida}, {Gomes}, {Gonz{\'a}lez-Marcos},
  {Gonz{\'a}lez-N{\'u}{\~n}ez}, {Gonz{\'a}lez-Vidal}, {Granvik}, {Guerrier},
  {Guillout}, {Guiraud}, {G{\'u}rpide}, {Guti{\'e}rrez-S{\'a}nchez}, {Guy},
  {Haigron}, {Hatzidimitriou}, {Haywood}, {Heiter}, {Helmi}, {Hobbs},
  {Hofmann}, {Holl}, {Holland}, {Hunt}, {Hypki}, {Icardi}, {Irwin}, {Jevardat
  de Fombelle}, {Jofr{\'e}}, {Jonker}, {Jorissen}, {Julbe}, {Karampelas},
  {Kochoska}, {Kohley}, {Kolenberg}, {Kontizas}, {Koposov}, {Kordopatis},
  {Koubsky}, {Kowalczyk}, {Krone-Martins}, {Kudryashova}, {Kull}, {Bachchan},
  {Lacoste-Seris}, {Lanza}, {Lavigne}, {Le Poncin-Lafitte}, {Lebreton},
  {Lebzelter}, {Leccia}, {Leclerc}, {Lecoeur-Taibi}, {Lemaitre}, {Lenhardt},
  {Leroux}, {Liao}, {Licata}, {Lindstr{\o}m}, {Lister}, {Livanou}, {Lobel},
  {L{\"o}ffler}, {L{\'o}pez}, {Lopez-Lozano}, {Lorenz}, {Loureiro},
  {MacDonald}, {Magalh{\~a}es Fernandes}, {Managau}, {Mann}, {Mantelet},
  {Marchal}, {Marchant}, {Marconi}, {Marie}, {Marinoni}, {Marrese},
  {Marschalk{\'o}}, {Marshall}, {Mart{\'\i}n-Fleitas}, {Martino}, {Mary},
  {Matijevi{\v{c}}}, {Mazeh}, {McMillan}, {Messina}, {Mestre}, {Michalik},
  {Millar}, {Miranda}, {Molina}, {Molinaro}, {Molinaro}, {Moln{\'a}r},
  {Moniez}, {Montegriffo}, {Monteiro}, {Mor}, {Mora}, {Morbidelli}, {Morel},
  {Morgenthaler}, {Morley}, {Morris}, {Mulone}, {Muraveva}, {Musella},
  {Narbonne}, {Nelemans}, {Nicastro}, {Noval}, {Ord{\'e}novic},
  {Ordieres-Mer{\'e}}, {Osborne}, {Pagani}, {Pagano}, {Pailler}, {Palacin},
  {Palaversa}, {Parsons}, {Paulsen}, {Pecoraro}, {Pedrosa}, {Pentik{\"a}inen},
  {Pereira}, {Pichon}, {Piersimoni}, {Pineau}, {Plachy}, {Plum}, {Poujoulet},
  {Pr{\v{s}}a}, {Pulone}, {Ragaini}, {Rago}, {Rambaux}, {Ramos-Lerate},
  {Ranalli}, {Rauw}, {Read}, {Regibo}, {Renk}, {Reyl{\'e}}, {Ribeiro},
  {Rimoldini}, {Ripepi}, {Riva}, {Rixon}, {Roelens}, {Romero-G{\'o}mez},
  {Rowell}, {Royer}, {Rudolph}, {Ruiz-Dern}, {Sadowski}, {Sagrist{\`a}
  Sell{\'e}s}, {Sahlmann}, {Salgado}, {Salguero}, {Sarasso}, {Savietto},
  {Schnorhk}, {Schultheis}, {Sciacca}, {Segol}, {Segovia}, {Segransan},
  {Serpell}, {Shih}, {Smareglia}, {Smart}, {Smith}, {Solano}, {Solitro},
  {Sordo}, {Soria Nieto}, {Souchay}, {Spagna}, {Spoto}, {Stampa}, {Steele},
  {Steidelm{\"u}ller}, {Stephenson}, {Stoev}, {Suess}, {S{\"u}veges}, {Surdej},
  {Szabados}, {Szegedi-Elek}, {Tapiador}, {Taris}, {Tauran}, {Taylor},
  {Teixeira}, {Terrett}, {Tingley}, {Trager}, {Turon}, {Ulla}, {Utrilla},
  {Valentini}, {van Elteren}, {Van Hemelryck}, {van Leeuwen}, {Varadi},
  {Vecchiato}, {Veljanoski}, {Via}, {Vicente}, {Vogt}, {Voss}, {Votruba},
  {Voutsinas}, {Walmsley}, {Weiler}, {Weingrill}, {Werner}, {Wevers},
  {Whitehead}, {Wyrzykowski}, {Yoldas}, {{\v{Z}}erjal}, {Zucker}, {Zurbach},
  {Zwitter}, {Alecu}, {Allen}, {Allende Prieto}, {Amorim},
  {Anglada-Escud{\'e}}, {Arsenijevic}, {Azaz}, {Balm}, {Beck}, {Bernstein},
  {Bigot}, {Bijaoui}, {Blasco}, {Bonfigli}, {Bono}, {Boudreault}, {Bressan},
  {Brown}, {Brunet}, {Bunclark}, {Buonanno}, {Butkevich}, {Carret}, {Carrion},
  {Chemin}, {Ch{\'e}reau}, {Corcione}, {Darmigny}, {de Boer}, {de Teodoro}, {de
  Zeeuw}, {Delle Luche}, {Domingues}, {Dubath}, {Fodor}, {Fr{\'e}zouls},
  {Fries}, {Fustes}, {Fyfe}, {Gallardo}, {Gallegos}, {Gardiol}, {Gebran},
  {Gomboc}, {G{\'o}mez}, {Grux}, {Gueguen}, {Heyrovsky}, {Hoar}, {Iannicola},
  {Isasi Parache}, {Janotto}, {Joliet}, {Jonckheere}, {Keil}, {Kim},
  {Klagyivik}, {Klar}, {Knude}, {Kochukhov}, {Kolka}, {Kos}, {Kutka}, {Lainey},
  {LeBouquin}, {Liu}, {Loreggia}, {Makarov}, {Marseille}, {Martayan},
  {Martinez-Rubi}, {Massart}, {Meynadier}, {Mignot}, {Munari}, {Nguyen},
  {Nordlander}, {Ocvirk}, {O'Flaherty}, {Olias Sanz}, {Ortiz}, {Osorio},
  {Oszkiewicz}, {Ouzounis}, {Palmer}, {Park}, {Pasquato}, {Peltzer}, {Peralta},
  {P{\'e}turaud}, {Pieniluoma}, {Pigozzi}, {Poels}, {Prat}, {Prod'homme},
  {Raison}, {Rebordao}, {Risquez}, {Rocca-Volmerange}, {Rosen}, {Ruiz-Fuertes},
  {Russo}, {Sembay}, {Serraller Vizcaino}, {Short}, {Siebert}, {Silva},
  {Sinachopoulos}, {Slezak}, {Soffel}, {Sosnowska}, {Strai{\v{z}}ys}, {ter
  Linden}, {Terrell}, {Theil}, {Tiede}, {Troisi}, {Tsalmantza}, {Tur},
  {Vaccari}, {Vachier}, {Valles}, {Van Hamme}, {Veltz}, {Virtanen}, {Wallut},
  {Wichmann}, {Wilkinson}, {Ziaeepour}, \& {Zschocke}}]{Gaia2016}
{Gaia Collaboration}, {Prusti}, T., {de Bruijne}, J.~H.~J., {et~al.} 2016,
  \aap, 595, A1

\bibitem[{{Golubov} {et~al.}(2013){Golubov}, {Just}, {Bienaym{\'e}},
  {Bland-Hawthorn}, {Gibson}, {Grebel}, {Munari}, {Navarro}, {Parker},
  {Seabroke}, {Reid}, {Siviero}, {Steinmetz}, {Williams}, {Watson}, \&
  {Zwitter}}]{Golubov2013}
{Golubov}, O., {Just}, A., {Bienaym{\'e}}, O., {et~al.} 2013, \aap, 557, A92

\bibitem[{{Gould}(1992)}]{Gould1992}
{Gould}, A. 1992, \apj, 392, 442

\bibitem[{{Gould}(1994{\natexlab{a}})}]{Gould1994b}
{Gould}, A. 1994{\natexlab{a}}, \apjl, 421, L75

\bibitem[{{Gould}(1994{\natexlab{b}})}]{Gould1994}
{Gould}, A. 1994{\natexlab{b}}, \apjl, 421, L71

\bibitem[{{Gould}(2000)}]{Gould2000}
{Gould}, A. 2000, \apj, 542, 785

\bibitem[{{Gould} {et~al.}(2009){Gould}, {Udalski}, {Monard}, {Horne}, {Dong},
  {Miyake}, {Sahu}, {Bennett}, {Wyrzykowski}, {Soszy{\'n}ski}, {Szyma{\'n}ski},
  {Kubiak}, {Pietrzy{\'n}ski}, {Szewczyk}, {Ulaczyk}, {OGLE Collaboration},
  {Allen}, {Christie}, {DePoy}, {Gaudi}, {Han}, {Lee}, {McCormick}, {Natusch},
  {Park}, {Pogge}, {{\ensuremath{\mu}}FUN Collaboration}, {Allan}, {Bode},
  {Bramich}, {Burgdorf}, {Dominik}, {Fraser}, {Kerins}, {Mottram}, {Snodgrass},
  {Steele}, {Street}, {Tsapras}, {RoboNet Collaboration}, {Abe}, {Bond},
  {Botzler}, {Fukui}, {Furusawa}, {Hearnshaw}, {Itow}, {Kamiya}, {Kilmartin},
  {Korpela}, {Lin}, {Ling}, {Masuda}, {Matsubara}, {Muraki}, {Nagaya},
  {Ohnishi}, {Okumura}, {Perrott}, {Rattenbury}, {Saito}, {Sako}, {Skuljan},
  {Sullivan}, {Sumi}, {Sweatman}, {Tristram}, {Yock}, {MOA Collaboration},
  {Albrow}, {Beaulieu}, {Coutures}, {Calitz}, {Caldwell}, {Fouque}, {Martin},
  {Williams}, \& {PLANET Collaboration}}]{Gould2009}
{Gould}, A., {Udalski}, A., {Monard}, B., {et~al.} 2009, \apjl, 698, L147

\bibitem[{{Gould} \& {Yee}(2012)}]{Gould2012}
{Gould}, A. \& {Yee}, J.~C. 2012, \apjl, 755, L17

\bibitem[{{Grether} \& {Lineweaver}(2006)}]{Grether2006}
{Grether}, D. \& {Lineweaver}, C.~H. 2006, \apj, 640, 1051

\bibitem[{{Han} \& {Chang}(2003)}]{Han2003}
{Han}, C. \& {Chang}, H.-Y. 2003, \mnras, 338, 637

\bibitem[{{Han} {et~al.}(2013){Han}, {Jung}, {Udalski}, {Sumi}, {Gaudi},
  {Gould}, {Bennett}, {Tsapras}, {Szyma{\'n}ski}, {Kubiak}, {Pietrzy{\'n}ski},
  {Soszy{\'n}ski}, {Skowron}, {Koz{\l}owski}, {Poleski}, {Ulaczyk},
  {Wyrzykowski}, {Pietrukowicz}, {OGLE Collaboration}, {Abe}, {Bond},
  {Botzler}, {Chote}, {Freeman}, {Fukui}, {Furusawa}, {Harris}, {Itow}, {Ling},
  {Masuda}, {Matsubara}, {Muraki}, {Ohnishi}, {Rattenbury}, {Saito},
  {Sullivan}, {Sweatman}, {Suzuki}, {Tristram}, {Wada}, {Yock}, {MOA
  Collaboration}, {Batista}, {Christie}, {Choi}, {DePoy}, {Dong}, {Hwang},
  {Kavka}, {Lee}, {Monard}, {Natusch}, {Ngan}, {Park}, {Pogge}, {Porritt},
  {Shin}, {Tan}, {Yee}, {{\ensuremath{\mu}}FUN Collaboration}, {Alsubai},
  {Bozza}, {Bramich}, {Browne}, {Dominik}, {Horne}, {Hundertmark}, {Ipatov},
  {Kains}, {Liebig}, {Snodgrass}, {Steele}, {Street}, \& {RoboNet
  Collaboration}}]{Han2013}
{Han}, C., {Jung}, Y.~K., {Udalski}, A., {et~al.} 2013, \apj, 778, 38

\bibitem[{{Irwin} {et~al.}(2010){Irwin}, {Buchhave}, {Berta}, {Charbonneau},
  {Latham}, {Burke}, {Esquerdo}, {Everett}, {Holman}, {Nutzman}, {Berlind},
  {Calkins}, {Falco}, {Winn}, {Johnson}, \& {Gazak}}]{Irwin2010}
{Irwin}, J., {Buchhave}, L., {Berta}, Z.~K., {et~al.} 2010, \apj, 718, 1353

\bibitem[{{Johnson} {et~al.}(2011){Johnson}, {Apps}, {Gazak}, {Crepp},
  {Crossfield}, {Howard}, {Marcy}, {Morton}, {Chubak}, \&
  {Isaacson}}]{johns2011}
{Johnson}, J.~A., {Apps}, K., {Gazak}, J.~Z., {et~al.} 2011, \apj, 730, 79

\bibitem[{{Johnson} {et~al.}(2020){Johnson}, {Penny}, {Gaudi}, {Kerins},
  {Rattenbury}, {Robin}, {Calchi Novati}, \& {Henderson}}]{Johnson2020}
{Johnson}, S.~A., {Penny}, M., {Gaudi}, B.~S., {et~al.} 2020, \aj, 160, 123

\bibitem[{{Kervella} {et~al.}(2004){Kervella}, {Th{\'e}venin}, {Di Folco}, \&
  {S{\'e}gransan}}]{Kervella2004}
{Kervella}, P., {Th{\'e}venin}, F., {Di Folco}, E., \& {S{\'e}gransan}, D.
  2004, \aap, 426, 297

\bibitem[{{Kim} {et~al.}(2021){Kim}, {Hwang}, {Gould}, {Yee}, {Ryu}, {Albrow},
  {Chung}, {Han}, {Kil Jung}, {Lee}, {Shin}, {Shvartzvald}, {Zang}, {Cha},
  {Kim}, {Kim}, {Lee}, {Lee}, {Park}, \& {Pogge}}]{Kim2021}
{Kim}, H.-W., {Hwang}, K.-H., {Gould}, A., {et~al.} 2021, \aj, 162, 15

\bibitem[{{Kim} {et~al.}(2016){Kim}, {Lee}, {Park}, {Kim}, {Cha}, {Lee}, {Han},
  {Chun}, \& {Yuk}}]{Kim2016}
{Kim}, S.-L., {Lee}, C.-U., {Park}, B.-G., {et~al.} 2016, Journal of Korean
  Astronomical Society, 49, 37

\bibitem[{{Kl{\"u}ter} {et~al.}(2020){Kl{\"u}ter}, {Bastian}, \&
  {Wambsganss}}]{Kluter2020}
{Kl{\"u}ter}, J., {Bastian}, U., \& {Wambsganss}, J. 2020, \aap, 640, A83

\bibitem[{{Koshimoto} {et~al.}(2021){Koshimoto}, {Baba}, \&
  {Bennett}}]{Koshimoto2021}
{Koshimoto}, N., {Baba}, J., \& {Bennett}, D.~P. 2021, \apj, 917, 78

\bibitem[{{Koshimoto} \& {Bennett}(2020)}]{Koshimoto2020}
{Koshimoto}, N. \& {Bennett}, D.~P. 2020, \aj, 160, 177

\bibitem[{{Lafreni{\`e}re} {et~al.}(2007){Lafreni{\`e}re}, {Doyon}, {Marois},
  {Nadeau}, {Oppenheimer}, {Roche}, {Rigaut}, {Graham}, {Jayawardhana},
  {Johnstone}, {Kalas}, {Macintosh}, \& {Racine}}]{lafren2007}
{Lafreni{\`e}re}, D., {Doyon}, R., {Marois}, C., {et~al.} 2007, \apj, 670, 1367

\bibitem[{{Lam} {et~al.}(2022){Lam}, {Lu}, {Udalski}, {Bond}, {Bennett},
  {Skowron}, {Mroz}, {Poleski}, {Sumi}, {Szymanski}, {Kozlowski},
  {Pietrukowicz}, {Soszynski}, {Ulaczyk}, {Wyrzykowski}, {Miyazaki}, {Suzuki},
  {Koshimoto}, {Rattenbury}, {Hosek}, {Abe}, {Barry}, {Bhattacharya}, {Fukui},
  {Fujii}, {Hirao}, {Itow}, {Kirikawa}, {Kondo}, {Matsubara}, {Matsumoto},
  {Muraki}, {Olmschenk}, {Ranc}, {Okamura}, {Satoh}, {Ishitani Silva}, {Toda},
  {Tristram}, {Vandorou}, {Yama}, {Abrams}, {Agarwal}, {Rose}, \&
  {Terry}}]{Lam2022}
{Lam}, C.~Y., {Lu}, J.~R., {Udalski}, A., {et~al.} 2022, arXiv e-prints,
  arXiv:2202.01903

\bibitem[{{Larson}(1992)}]{Larson1992}
{Larson}, R.~B. 1992, \mnras, 256, 641

\bibitem[{{Ma} \& {Zhu}(2021)}]{Ma2021}
{Ma}, X. \& {Zhu}, W. 2021, arXiv e-prints, arXiv:2111.11059

\bibitem[{{Mao} \& {Di Stefano}(1995)}]{Mao1995}
{Mao}, S. \& {Di Stefano}, R. 1995, \apj, 440, 22

\bibitem[{{Marcy} \& {Butler}(2000)}]{Marcy2000}
{Marcy}, G.~W. \& {Butler}, R.~P. 2000, \pasp, 112, 137

\bibitem[{{McLean} {et~al.}(2003){McLean}, {McGovern}, {Burgasser},
  {Kirkpatrick}, {Prato}, \& {Kim}}]{Mclean2003}
{McLean}, I.~S., {McGovern}, M.~R., {Burgasser}, A.~J., {et~al.} 2003, \apj,
  596, 561

\bibitem[{{Meisner} {et~al.}(2020){Meisner}, {Faherty}, {Kirkpatrick},
  {Schneider}, {Caselden}, {Gagn{\'e}}, {Kuchner}, {Burgasser}, {Casewell},
  {Debes}, {Artigau}, {Bardalez Gagliuffi}, {Logsdon}, {Kiman}, {Allers},
  {Hsu}, {Wisniewski}, {Allen}, {Beaulieu}, {Colin}, {Durantini Luca},
  {Goodman}, {Gramaize}, {Hamlet}, {Hinckley}, {Kiwy}, {Martin}, {Pendrill},
  {Rothermich}, {Sainio}, {Sch{\"u}mann}, {Andersen}, {Tanner}, {Thakur},
  {Th{\'e}venot}, {Walla}, {W{\k{e}}dracki}, {Aganze}, {Gerasimov}, {Theissen},
  \& {Backyard Worlds: Planet 9 Collaboration}}]{Meisner2020}
{Meisner}, A.~M., {Faherty}, J.~K., {Kirkpatrick}, J.~D., {et~al.} 2020, \apj,
  899, 123

\bibitem[{{Miret-Roig} {et~al.}(2021){Miret-Roig}, {Bouy}, {Raymond}, {Tamura},
  {Bertin}, {Barrado}, {Olivares}, {Galli}, {Cuillandre}, {Sarro}, {Berihuete},
  \& {Hu{\'e}lamo}}]{Miret-Roig2021}
{Miret-Roig}, N., {Bouy}, H., {Raymond}, S.~N., {et~al.} 2021, Nature Astronomy
  [\eprint[arXiv]{2112.11999}]

\bibitem[{{Miyake} {et~al.}(2011){Miyake}, {Sumi}, {Dong}, {Street}, {Mancini},
  {Gould}, {Bennett}, {Tsapras}, {Yee}, {Albrow}, {Bond}, {Fouqu{\'e}},
  {Browne}, {Han}, {Snodgrass}, {Finet}, {Furusawa}, {Harps{\o}e}, {Allen},
  {Hundertmark}, {Freeman}, {Suzuki}, {Abe}, {Botzler}, {Douchin}, {Fukui},
  {Hayashi}, {Hearnshaw}, {Hosaka}, {Itow}, {Kamiya}, {Kilmartin}, {Korpela},
  {Lin}, {Ling}, {Makita}, {Masuda}, {Matsubara}, {Muraki}, {Nagayama},
  {Nishimoto}, {Ohnishi}, {Perrott}, {Rattenbury}, {Saito}, {Skuljan},
  {Sullivan}, {Sweatman}, {Tristram}, {Wada}, {Yock}, {MOA Collaboration},
  {Bolt}, {Bos}, {Christie}, {DePoy}, {Drummond}, {Gal-Yam}, {Gaudi},
  {Gorbikov}, {Higgins}, {Hwang}, {Janczak}, {Kaspi}, {Lee}, {Koo},
  {Koz{\l}owski}, {Lee}, {Mallia}, {Maury}, {Maoz}, {McCormick}, {Monard},
  {Moorhouse}, {Mu{\~n}oz}, {Natusch}, {Ofek}, {Pogge}, {Polishook},
  {Santallo}, {Shporer}, {Spector}, {Thornley}, {{\ensuremath{\mu}}FUN
  Collaboration}, {Allan}, {Bramich}, {Horne}, {Kains}, {Steele}, {RoboNet
  Collaboration}, {Bozza}, {Burgdorf}, {Calchi Novati}, {Dominik}, {Dreizler},
  {Glitrup}, {Hessman}, {Hinse}, {J{\o}rgensen}, {Liebig}, {Maier},
  {Mathiasen}, {Rahvar}, {Ricci}, {Scarpetta}, {Skottfelt}, {Southworth},
  {Surdej}, {Wambsganss}, {Zimmer}, {MiNDSTEp Consortium}, {Batista},
  {Beaulieu}, {Brillant}, {Cassan}, {Cole}, {Corrales}, {Coutures}, {Dieters},
  {Greenhill}, {Kubas}, {Menzies}, \& {PLANET Collaboration}}]{Miyake2011}
{Miyake}, N., {Sumi}, T., {Dong}, S., {et~al.} 2011, \apj, 728, 120

\bibitem[{{Miyazaki} {et~al.}(2021){Miyazaki}, {Johnson}, {Sumi}, {Penny},
  {Koshimoto}, \& {Yamawaki}}]{Miyazaki2021}
{Miyazaki}, S., {Johnson}, S.~A., {Sumi}, T., {et~al.} 2021, \aj, 161, 84

\bibitem[{{Molli{\`e}re} \& {Mordasini}(2012)}]{molli2012}
{Molli{\`e}re}, P. \& {Mordasini}, C. 2012, \aap, 547, A105

\bibitem[{{Mr{\'o}z} {et~al.}(2020){Mr{\'o}z}, {Poleski}, {Gould}, {Udalski},
  {Sumi}, {Szyma{\'n}ski}, {Soszy{\'n}ski}, {Pietrukowicz}, {Koz{\l}owski},
  {Skowron}, {Ulaczyk}, {OGLE Collaboration}, {Albrow}, {Chung}, {Han},
  {Hwang}, {Jung}, {Kim}, {Ryu}, {Shin}, {Shvartzvald}, {Yee}, {Zang}, {Cha},
  {Kim}, {Kim}, {Lee}, {Lee}, {Lee}, {Park}, {Pogge}, \& {KMT
  Collaboration}}]{Mroz2020}
{Mr{\'o}z}, P., {Poleski}, R., {Gould}, A., {et~al.} 2020, \apjl, 903, L11

\bibitem[{{Mr{\'o}z} {et~al.}(2017){Mr{\'o}z}, {Udalski}, {Skowron}, {Poleski},
  {Koz{\l}owski}, {Szyma{\'n}ski}, {Soszy{\'n}ski}, {Wyrzykowski},
  {Pietrukowicz}, {Ulaczyk}, {Skowron}, \& {Pawlak}}]{Mroz2017}
{Mr{\'o}z}, P., {Udalski}, A., {Skowron}, J., {et~al.} 2017, \nat, 548, 183

\bibitem[{{Mr{\'o}z} {et~al.}(2019){Mr{\'o}z}, {Udalski}, {Skowron},
  {Szyma{\'n}ski}, {Soszy{\'n}ski}, {Wyrzykowski}, {Pietrukowicz},
  {Koz{\l}owski}, {Poleski}, {Ulaczyk}, {Rybicki}, \& {Iwanek}}]{Mroz2019}
{Mr{\'o}z}, P., {Udalski}, A., {Skowron}, J., {et~al.} 2019, \apjs, 244, 29

\bibitem[{{Nataf} {et~al.}(2013){Nataf}, {Gould}, {Fouqu{\'e}}, {Gonzalez},
  {Johnson}, {Skowron}, {Udalski}, {Szyma{\'n}ski}, {Kubiak},
  {Pietrzy{\'n}ski}, {Soszy{\'n}ski}, {Ulaczyk}, {Wyrzykowski}, \&
  {Poleski}}]{Nataf2013}
{Nataf}, D.~M., {Gould}, A., {Fouqu{\'e}}, P., {et~al.} 2013, \apj, 769, 88

\bibitem[{{Offner} {et~al.}(2010){Offner}, {Kratter}, {Matzner}, {Krumholz}, \&
  {Klein}}]{Stella2010}
{Offner}, S. S.~R., {Kratter}, K.~M., {Matzner}, C.~D., {Krumholz}, M.~R., \&
  {Klein}, R.~I. 2010, \apj, 725, 1485

\bibitem[{{Padoan} {et~al.}(2005){Padoan}, {Kritsuk}, {Michael}, {Norman}, \&
  {Nordlund}}]{padoan2004}
{Padoan}, P., {Kritsuk}, A., {Michael}, {Norman}, L., \& {Nordlund}, {\r{A}}.
  2005, \memsai, 76, 187

\bibitem[{{Penny} {et~al.}(2019){Penny}, {Gaudi}, {Kerins}, {Rattenbury},
  {Mao}, {Robin}, \& {Calchi Novati}}]{Penny2019}
{Penny}, M.~T., {Gaudi}, B.~S., {Kerins}, E., {et~al.} 2019, \apjs, 241, 3

\bibitem[{{Poindexter} {et~al.}(2005){Poindexter}, {Afonso}, {Bennett},
  {Glicenstein}, {Gould}, {Szyma{\'n}ski}, \& {Udalski}}]{Poindexter2005}
{Poindexter}, S., {Afonso}, C., {Bennett}, D.~P., {et~al.} 2005, \apj, 633, 914

\bibitem[{{Ranc} {et~al.}(2015){Ranc}, {Cassan}, {Albrow}, {Kubas}, {Bond},
  {Batista}, {Beaulieu}, {Bennett}, {Dominik}, {Dong}, {Fouqu{\'e}}, {Gould},
  {Greenhill}, {J{\o}rgensen}, {Kains}, {Menzies}, {Sumi}, {Bachelet},
  {Coutures}, {Dieters}, {Dominis Prester}, {Donatowicz}, {Gaudi}, {Han},
  {Hundertmark}, {Horne}, {Kane}, {Lee}, {Marquette}, {Park}, {Pollard},
  {Sahu}, {Street}, {Tsapras}, {Wambsganss}, {Williams}, {Zub}, {Abe}, {Fukui},
  {Itow}, {Masuda}, {Matsubara}, {Muraki}, {Ohnishi}, {Rattenbury}, {Saito},
  {Sullivan}, {Sweatman}, {Tristram}, {Yock}, \& {Yonehara}}]{Ranc2015}
{Ranc}, C., {Cassan}, A., {Albrow}, M.~D., {et~al.} 2015, \aap, 580, A125

\bibitem[{{Refsdal}(1966)}]{Refsdal1966}
{Refsdal}, S. 1966, \mnras, 134, 315

\bibitem[{{Ryan} \& {Reid}(2016)}]{Ryan2016}
{Ryan}, R.~E., J. \& {Reid}, I.~N. 2016, \aj, 151, 92

\bibitem[{{Ryu} {et~al.}(2021){Ryu}, {Hwang}, {Gould}, {Yee}, {Albrow},
  {Chung}, {Han}, {Jung}, {Kim}, {Shin}, {Shvartzvald}, {Zang}, {Cha}, {Kim},
  {Kim}, {Lee}, {Lee}, {Lee}, {Park}, \& {Pogge}}]{Ryu2021}
{Ryu}, Y.-H., {Hwang}, K.-H., {Gould}, A., {et~al.} 2021, \aj, 162, 96

\bibitem[{{Sahlmann} {et~al.}(2011){Sahlmann}, {S{\'e}gransan}, {Queloz},
  {Udry}, {Santos}, {Marmier}, {Mayor}, {Naef}, {Pepe}, \&
  {Zucker}}]{sahlmann2011}
{Sahlmann}, J., {S{\'e}gransan}, D., {Queloz}, D., {et~al.} 2011, \aap, 525,
  A95

\bibitem[{{Sahu} {et~al.}(2022){Sahu}, {Anderson}, {Casertano}, {Bond},
  {Udalski}, {Dominik}, {Calamida}, {Bellini}, {Brown}, {Rejkuba}, {Bajaj},
  {Kains}, {Ferguson}, {Fryer}, {Yock}, {Mroz}, {Kozlowski}, {Pietrukowicz},
  {Poleski}, {Skowron}, {Soszynski}, {Szymanski}, {Ulaczyk}, {Wyrzykowski},
  {Beaulieu}, {Marquette}, {Cole}, {Hill}, {Dieters}, {Coutures},
  {Dominis-Prester}, {Bachelet}, {Menzies}, {Albrow}, {Pollard}, {Gould},
  {Yee}, {Allen}, {de Almeida}, {Christie}, {Drummond}, {Gal-Yam}, {Gorbikov},
  {Jablonski}, {Lee}, {Maoz}, {Manulis}, {McCormick}, {Natusch}, {Pogge},
  {Shvartzvald}, {Jorgensen}, {Alsubai}, {Andersen}, {Bozza}, {Calchi Novati},
  {Hinse}, {Hundertmark}, {Husser}, {Kerins}, {Longa-Pena}, {Mancini}, {Penny},
  {Rahvar}, {Ricci}, {Sajadian}, {Skottfelt}, {Snodgrass}, {Southworth},
  {Tregloan-Reed}, {Wambsganss}, {Wertz}, {Tsapras}, {Street}, {Bramich},
  {Horne}, \& {Steele}}]{Sahu2022}
{Sahu}, K.~C., {Anderson}, J., {Casertano}, S., {et~al.} 2022, arXiv e-prints,
  arXiv:2201.13296

\bibitem[{{Schechter} {et~al.}(1993){Schechter}, {Mateo}, \&
  {Saha}}]{Schechter1993}
{Schechter}, P.~L., {Mateo}, M., \& {Saha}, A. 1993, \pasp, 105, 1342

\bibitem[{{Shin} {et~al.}(2012){Shin}, {Han}, {Gould}, {Udalski}, {Sumi},
  {Dominik}, {Beaulieu}, {Tsapras}, {Bozza}, {Szyma{\'n}ski}, {Kubiak},
  {Soszy{\'n}ski}, {Pietrzy{\'n}ski}, {Poleski}, {Ulaczyk}, {Pietrukowicz},
  {Koz{\l}owski}, {Skowron}, {Wyrzykowski}, {OGLE Collaboration}, {Abe},
  {Bennett}, {Bond}, {Botzler}, {Freeman}, {Fukui}, {Furusawa}, {Hayashi},
  {Hearnshaw}, {Hosaka}, {Itow}, {Kamiya}, {Kilmartin}, {Kobara}, {Korpela},
  {Lin}, {Ling}, {Makita}, {Masuda}, {Matsubara}, {Miyake}, {Muraki}, {Nagaya},
  {Nishimoto}, {Ohnishi}, {Okumura}, {Omori}, {Perrott}, {Rattenbury}, {Saito},
  {Skuljan}, {Sullivan}, {Suzuki}, {Sweatman}, {Tristram}, {Wada}, {Yock}, {MOA
  Collaboration}, {Christie}, {Depoy}, {Dong}, {Gal-Yam}, {Gaudi}, {Hung},
  {Janczak}, {Kaspi}, {Maoz}, {McCormick}, {McGregor}, {Moorhouse},
  {Mu{\~n}oz}, {Natusch}, {Nelson}, {Pogge}, {Tan}, {Polishook}, {Shvartzvald},
  {Shporer}, {Thornley}, {Malamud}, {Yee}, {Choi}, {Jung}, {Park}, {Lee},
  {Park}, {Koo}, {{\ensuremath{\mu}}FUN Collaboration}, {Bajek}, {Bramich},
  {Browne}, {Horne}, {Ipatov}, {Snodgrass}, {Steele}, {Street}, {Alsubai},
  {Burgdorf}, {Calchi Novati}, {Dodds}, {Dreizler}, {Fang}, {Grundahl}, {Gu},
  {Hardis}, {Harps{\o}e}, {Hinse}, {Hundertmark}, {Jessen-Hansen},
  {J{\o}rgensen}, {Kains}, {Kerins}, {Liebig}, {Lund}, {Lundkvist}, {Mancini},
  {Mathiasen}, {Hornstrup}, {Penny}, {Proft}, {Rahvar}, {Ricci}, {Scarpetta},
  {Skottfelt}, {Southworth}, {Surdej}, {Tregloan-Reed}, {Wertz}, {Zimmer},
  {Albrow}, {Batista}, {Brillant}, {Caldwell}, {Calitz}, {Cassan}, {Cole},
  {Cook}, {Corrales}, {Coutures}, {Dieters}, {Dominis Prester}, {Donatowicz},
  {Fouqu{\'e}}, {Greenhill}, {Hill}, {Hoffman}, {Kane}, {Kubas}, {Marquette},
  {Martin}, {Meintjes}, {Menzies}, {Pollard}, {Sahu}, {Wambsganss}, {Williams},
  {Vinter}, \& {Zub}}]{Shin2012}
{Shin}, I.~G., {Han}, C., {Gould}, A., {et~al.} 2012, \apj, 760, 116

\bibitem[{{Shin} {et~al.}(2018){Shin}, {Udalski}, {Yee}, {Calchi Novati},
  {Christie}, {Poleski}, {Mr{\'o}z}, {Skowron}, {Szyma{\'n}ski},
  {Soszy{\'n}ski}, {Pietrukowicz}, {Koz{\l}owski}, {Ulaczyk}, {Pawlak}, {OGLE
  Collaboration}, {Natusch}, {Pogge}, {{\ensuremath{\mu}}FUN Collaboration},
  {Gould}, {Han}, {Albrow}, {Chung}, {Hwang}, {Ryu}, {Jung}, {Zhu}, {Lee},
  {Cha}, {Kim}, {Kim}, {Kim}, {Lee}, {Lee}, {Park}, {KMTNet Group}, {Beichman},
  {Bryden}, {Carey}, {Gaudi}, {Henderson}, {Shvartzvald}, \& {Spitzer
  Team}}]{Shin2018}
{Shin}, I.~G., {Udalski}, A., {Yee}, J.~C., {et~al.} 2018, \apj, 863, 23

\bibitem[{{Shin} {et~al.}(2022){Shin}, {Yee}, {Hwang}, {Gould}, {Udalski},
  {Bond}, {Albrow}, {Chung}, {Han}, {Jung}, {Kim}, {Ryu}, {Shvartzvald},
  {Zang}, {Cha}, {Kim}, {Kim}, {Lee}, {Lee}, {Lee}, {Park}, {Pogge},
  {Mr{\'o}z}, {Szyma{\'n}ski}, {Skowron}, {Poleski}, {Soszy{\'n}ski},
  {Pietrukowicz}, {Koz{\l}owski}, {Ulaczyk}, {Beichman}, {Bryden}, {Calchi
  Novati}, {Carey}, {Gaudi}, {Henderson}, {Zhu}, {Abe}, {Barry}, {Bennett},
  {Bhattacharya}, {Fujii}, {Fukui}, {Hirao}, {Itow}, {Kirikawa}, {Koshimoto},
  {Kondo}, {Matsubara}, {Matsumoto}, {Miyazaki}, {Muraki}, {Olmschenk},
  {Okamura}, {Ranc}, {Rattenbury}, {Satoh}, {Ishitani Silva}, {Sumi}, {Suzuki},
  {Toda}, {Tristram}, {Vandorou}, \& {Yama}}]{Shin2022}
{Shin}, I.-G., {Yee}, J.~C., {Hwang}, K.-H., {et~al.} 2022, arXiv e-prints,
  arXiv:2201.04312

\bibitem[{{Shvartzvald} {et~al.}(2016){Shvartzvald}, {Li}, {Udalski}, {Gould},
  {Sumi}, {Street}, {Calchi Novati}, {Hundertmark}, {Bozza}, {Beichman},
  {Bryden}, {Carey}, {Drummond}, {Fausnaugh}, {Gaudi}, {Henderson}, {Tan},
  {Wibking}, {Pogge}, {Yee}, {Zhu}, {(Spitzer Team}, {Tsapras}, {Bachelet},
  {Dominik}, {Bramich}, {Cassan}, {Figuera Jaimes}, {Horne}, {Ranc}, {Schmidt},
  {Snodgrass}, {Wambsganss}, {Steele}, {Menzies}, {Mao}, {(RoboNet}, {Poleski},
  {Pawlak}, {Szyma{\'n}ski}, {Skowron}, {Mr{\'o}z}, {Koz{\l}owski},
  {Wyrzykowski}, {Pietrukowicz}, {Soszy{\'n}ski}, {Ulaczyk}, {(OGLE Group},
  {Abe}, {Asakura}, {Barry}, {Bennett}, {Bhattacharya}, {Bond}, {Freeman},
  {Hirao}, {Itow}, {Koshimoto}, {Li}, {Ling}, {Masuda}, {Fukui}, {Matsubara},
  {Muraki}, {Nagakane}, {Nishioka}, {Ohnishi}, {Oyokawa}, {Rattenbury},
  {Saito}, {Sharan}, {Sullivan}, {Suzuki}, {Tristram}, {Yonehara}, {(MOA
  Group}, {J{\o}rgensen}, {Burgdorf}, {Ciceri}, {D'Ago}, {Evans}, {Hinse},
  {Kains}, {Kerins}, {Korhonen}, {Mancini}, {Popovas}, {Rabus}, {Rahvar},
  {Scarpetta}, {Skottfelt}, {Southworth}, {Peixinho}, {Verma}, {(MiNDSTEp},
  {Sbarufatti}, {Kennea}, {Gehrels}, \& {(Swift}}]{Shvartzvald2016}
{Shvartzvald}, Y., {Li}, Z., {Udalski}, A., {et~al.} 2016, \apj, 831, 183

\bibitem[{{Skottfelt} {et~al.}(2015){Skottfelt}, {Bramich}, {Hundertmark},
  {J{\o}rgensen}, {Michaelsen}, {Kj{\ae}rgaard}, {Southworth}, {S{\o}rensen},
  {Andersen}, {Andersen}, {Christensen-Dalsgaard}, {Frandsen}, {Grundahl},
  {Harps{\o}e}, {Kjeldsen}, \& {Pall{\'e}}}]{Skottfelt2015}
{Skottfelt}, J., {Bramich}, D.~M., {Hundertmark}, M., {et~al.} 2015, \aap, 574,
  A54

\bibitem[{{Skowron} {et~al.}(2011){Skowron}, {Udalski}, {Gould}, {Dong},
  {Monard}, {Han}, {Nelson}, {McCormick}, {Moorhouse}, {Thornley}, {Maury},
  {Bramich}, {Greenhill}, {Koz{\l}owski}, {Bond}, {Poleski}, {Wyrzykowski},
  {Ulaczyk}, {Kubiak}, {Szyma{\'n}ski}, {Pietrzy{\'n}ski}, {Soszy{\'n}ski},
  {OGLE Collaboration}, {Gaudi}, {Yee}, {Hung}, {Pogge}, {DePoy}, {Lee},
  {Park}, {Allen}, {Mallia}, {Drummond}, {Bolt}, {{\ensuremath{\mu}}FUN
  Collaboration}, {Allan}, {Browne}, {Clay}, {Dominik}, {Fraser}, {Horne},
  {Kains}, {Mottram}, {Snodgrass}, {Steele}, {Street}, {Tsapras}, {RoboNet
  Collaboration}, {Abe}, {Bennett}, {Botzler}, {Douchin}, {Freeman}, {Fukui},
  {Furusawa}, {Hayashi}, {Hearnshaw}, {Hosaka}, {Itow}, {Kamiya}, {Kilmartin},
  {Korpela}, {Lin}, {Ling}, {Makita}, {Masuda}, {Matsubara}, {Muraki},
  {Nagayama}, {Miyake}, {Nishimoto}, {Ohnishi}, {Perrott}, {Rattenbury},
  {Saito}, {Skuljan}, {Sullivan}, {Sumi}, {Suzuki}, {Sweatman}, {Tristram},
  {Wada}, {Yock}, {MOA Collaboration}, {Beaulieu}, {Fouqu{\'e}}, {Albrow},
  {Batista}, {Brillant}, {Caldwell}, {Cassan}, {Cole}, {Cook}, {Coutures},
  {Dieters}, {Dominis Prester}, {Donatowicz}, {Kane}, {Kubas}, {Marquette},
  {Martin}, {Menzies}, {Sahu}, {Wambsganss}, {Williams}, {Zub}, \& {PLANET
  Collaboration}}]{Skowron2011}
{Skowron}, J., {Udalski}, A., {Gould}, A., {et~al.} 2011, \apj, 738, 87

\bibitem[{{Smith} {et~al.}(2003){Smith}, {Mao}, \& {Paczy{\'n}ski}}]{Smith2003}
{Smith}, M.~C., {Mao}, S., \& {Paczy{\'n}ski}, B. 2003, \mnras, 339, 925

\bibitem[{{Spezzi} {et~al.}(2011){Spezzi}, {Beccari}, {De Marchi}, {Young},
  {Paresce}, {Dopita}, {Andersen}, {Panagia}, {Balick}, {Bond}, {Calzetti},
  {Carollo}, {Disney}, {Frogel}, {Hall}, {Holtzman}, {Kimble}, {McCarthy},
  {O'Connell}, {Ryan}, {Saha}, {Silk}, {Trauger}, {Walker}, {Whitmore}, \&
  {Windhorst}}]{spezzi2011}
{Spezzi}, L., {Beccari}, G., {De Marchi}, G., {et~al.} 2011, \apj, 731, 1

\bibitem[{{Spiegel} {et~al.}(2011){Spiegel}, {Burrows}, \&
  {Milsom}}]{Spiegel2011}
{Spiegel}, D.~S., {Burrows}, A., \& {Milsom}, J.~A. 2011, \apj, 727, 57

\bibitem[{{Sumi} {et~al.}(2003){Sumi}, {Abe}, {Bond}, {Dodd}, {Hearnshaw},
  {Honda}, {Honma}, {Kan-ya}, {Kilmartin}, {Masuda}, {Matsubara}, {Muraki},
  {Nakamura}, {Nishi}, {Noda}, {Ohnishi}, {Petterson}, {Rattenbury}, {Reid},
  {Saito}, {Saito}, {Sato}, {Sekiguchi}, {Skuljan}, {Sullivan}, {Takeuti},
  {Tristram}, {Wilkinson}, {Yanagisawa}, \& {Yock}}]{Sumi2003}
{Sumi}, T., {Abe}, F., {Bond}, I.~A., {et~al.} 2003, \apj, 591, 204

\bibitem[{{Tomaney} \& {Crotts}(1996)}]{Tomaney1996}
{Tomaney}, A.~B. \& {Crotts}, A. P.~S. 1996, \aj, 112, 2872

\bibitem[{{Trippe} {et~al.}(2010){Trippe}, {Davies}, {Eisenhauer}, {F{\"o}rster
  Schreiber}, {Fritz}, \& {Genzel}}]{Trippe2010}
{Trippe}, S., {Davies}, R., {Eisenhauer}, F., {et~al.} 2010, \mnras, 402, 1126

\bibitem[{{Udalski} {et~al.}(2015){Udalski}, {Szyma{\'n}ski}, \&
  {Szyma{\'n}ski}}]{Udalski2015}
{Udalski}, A., {Szyma{\'n}ski}, M.~K., \& {Szyma{\'n}ski}, G. 2015, \actaa, 65,
  1

\bibitem[{{Vedantham} {et~al.}(2020){Vedantham}, {Callingham}, {Shimwell},
  {Dupuy}, {Best}, {Liu}, {Zhang}, {De}, {Lamy}, {Zarka}, {R{\"o}ttgering}, \&
  {Shulevski}}]{HKveda2020}
{Vedantham}, H.~K., {Callingham}, J.~R., {Shimwell}, T.~W., {et~al.} 2020,
  \apjl, 903, L33

\bibitem[{{Whitworth} {et~al.}(2007){Whitworth}, {Bate}, {Nordlund},
  {Reipurth}, \& {Zinnecker}}]{Whitworth2007}
{Whitworth}, A., {Bate}, M.~R., {Nordlund}, {\r{A}}., {Reipurth}, B., \&
  {Zinnecker}, H. 2007, in Protostars and Planets V, ed. B.~{Reipurth},
  D.~{Jewitt}, \& K.~{Keil}, 459

\bibitem[{{Wozniak}(2000)}]{Wozniak2000}
{Wozniak}, P.~R. 2000, \actaa, 50, 421

\bibitem[{{Wyrzykowski} \& {Mandel}(2020)}]{Wyrzykowski2020}
{Wyrzykowski}, {\L}. \& {Mandel}, I. 2020, \aap, 636, A20

\bibitem[{{Yee} {et~al.}(2015){Yee}, {Gould}, {Beichman}, {Calchi Novati},
  {Carey}, {Gaudi}, {Henderson}, {Nataf}, {Penny}, {Shvartzvald}, \&
  {Zhu}}]{Yee2015}
{Yee}, J.~C., {Gould}, A., {Beichman}, C., {et~al.} 2015, \apj, 810, 155

\bibitem[{{Yee} {et~al.}(2013){Yee}, {Hung}, {Bond}, {Allen}, {Monard},
  {Albrow}, {Fouqu{\'e}}, {Dominik}, {Tsapras}, {Udalski}, {Gould}, {Zellem},
  {Bos}, {Christie}, {DePoy}, {Dong}, {Drummond}, {Gaudi}, {Gorbikov}, {Han},
  {Kaspi}, {Klein}, {Lee}, {Maoz}, {McCormick}, {Moorhouse}, {Natusch}, {Nola},
  {Park}, {Pogge}, {Polishook}, {Shporer}, {Shvartzvald}, {Skowron},
  {Thornley}, {{\ensuremath{\mu}}FUN Collaboration}, {Abe}, {Bennett},
  {Botzler}, {Chote}, {Freeman}, {Fukui}, {Furusawa}, {Harris}, {Itow}, {Ling},
  {Masuda}, {Matsubara}, {Miyake}, {Ohnishi}, {Rattenbury}, {Saito},
  {Sullivan}, {Sumi}, {Suzuki}, {Sweatman}, {Tristram}, {Wada}, {Yock}, {MOA
  Collaboration}, {Szyma{\'n}ski}, {Soszy{\'n}ski}, {Kubiak}, {Poleski},
  {Ulaczyk}, {Pietrzy{\'n}ski}, {Wyrzykowski}, {OGLE Collaboration},
  {Bachelet}, {Batista}, {Beatty}, {Beaulieu}, {Bennett}, {Bowens-Rubin},
  {Brillant}, {Caldwell}, {Cassan}, {Cole}, {Corrales}, {Coutures}, {Dieters},
  {Dominis Prester}, {Donatowicz}, {Greenhill}, {Henderson}, {Kubas},
  {Marquette}, {Martin}, {Menzies}, {Shappee}, {Williams}, {Wouters}, {van
  Saders}, {Zub}, {PLANET Collaboration}, {Street}, {Horne}, {Bramich},
  {Steele}, {RoboNet Collaboration}, {Alsubai}, {Bozza}, {Browne}, {Burgdorf},
  {Calchi Novati}, {Dodds}, {Finet}, {Gerner}, {Hardis}, {Harps{\o}e},
  {Hessman}, {Hinse}, {Hundertmark}, {J{\o}rgensen}, {Kains}, {Kerins},
  {Liebig}, {Mancini}, {Mathiasen}, {Penny}, {Proft}, {Rahvar}, {Ricci},
  {Sahu}, {Scarpetta}, {Sch{\"a}fer}, {Sch{\"o}nebeck}, {Snodgrass},
  {Southworth}, {Surdej}, {Wambsganss}, \& {MiNDSTEp Consortium}}]{Yee2013}
{Yee}, J.~C., {Hung}, L.~W., {Bond}, I.~A., {et~al.} 2013, \apj, 769, 77

\bibitem[{{Yoo} {et~al.}(2004){Yoo}, {DePoy}, {Gal-Yam}, {Gaudi}, {Gould},
  {Han}, {Lipkin}, {Maoz}, {Ofek}, {Park}, {Pogge}, {Mu-Fun Collaboration},
  {Udalski}, {Soszy{\'n}ski}, {Wyrzykowski}, {Kubiak}, {Szyma{\'n}ski},
  {Pietrzy{\'n}ski}, {Szewczyk}, {{\.Z}ebru{\'n}}, \& {OGLE
  Collaboration}}]{yoo2004}
{Yoo}, J., {DePoy}, D.~L., {Gal-Yam}, A., {et~al.} 2004, \apj, 603, 139

\bibitem[{{Zang} {et~al.}(2018){Zang}, {Penny}, {Zhu}, {Mao}, {Fouqu{\'e}},
  {Udalski}, {Hwang}, {Wang}, {Huang}, {Boyajian}, \& {Barentsen}}]{Zang2018}
{Zang}, W., {Penny}, M.~T., {Zhu}, W., {et~al.} 2018, \pasp, 130, 104401

\end{thebibliography}
%

%

\end{document}